\documentclass[12pt, oneside, a4paper]{article}
\usepackage{mcite}
\usepackage{cite}
\usepackage{latexsym}
\usepackage{amssymb}
\usepackage{amsmath}
\usepackage{amsthm}
\usepackage[applemac]{inputenc}
\newcommand{\be}{\begin{eqnarray}}
            \newcommand{\ee}{\end{eqnarray}}
           \newcommand{\eel}[1]{\label{#1}\end{eqnarray}}
\newcommand{\e}[1]{\label{eq:#1}\end{eqnarray}}
     \newcommand{\eg}{{\em e.g.\ }}
            \newcommand{\ie}{{\em i.e.\ }}
            \newcommand{\ga}{{\gamma}}
 
            \newcommand{\la}{{\lambda}}

\newcommand{\del}{{\delta}}

\newcommand{\dy}{{\dot{y}}}

           \newcommand{\ra}{{\rightarrow}}

\newcommand{\dox}{{\dot x}}

\newcommand{\cL}{{\cal L}}

            \newcommand{\beq}{\begin{quote}}
            \newcommand{\eq}{\end{quote}}
            
            \newcommand{\al}{\alpha}
            \newcommand{\ben}{\begin{enumerate}}
            \newcommand{\een}{\end{enumerate}}
            \newcommand{\bit}{\begin{itemize}}
            \newcommand{\ei}{\end{itemize}}
        \newcommand{\nn}{\nonumber}
            \newcommand{\rl}[1]{(\ref{eq:#1})}
            \newcommand{\edfl}[1]{\Label{#1}\end{df}}

\newcommand{\hb}{{\cal i}}

\newcommand{\cF}{{\cal F}}
\newcommand{\cR}{{\cal R}}

\newcommand{\dif}{{\partial}}
\newcommand{\half}{\frac{1}{2}}

\begin{document}
\begin{titlepage}
%\today
\vspace*{5 mm}
\vspace*{20mm}
\begin{center}
{\LARGE\bf Conformal theories including conformal gravity as gauge theories on the hypercone}\end{center}
%\begin{center}{\LARGE\bf Manifestly conformal theories}\end{center}
\vspace*{3 mm}
\begin{center}
\vspace*{3 mm}

\begin{center}Pär Arvidsson\footnote{E-mail:
par.arvidsson@chalmers.se} and Robert Marnelius\footnote{E-mail: 
tferm@fy.chalmers.se}
 \\ \vspace*{7 mm} {\sl
Department of Fundamental Physics\\ Chalmers University of Technology\\
S-412 96  G\"{o}teborg, Sweden}\end{center}
\vspace*{25 mm}
\begin{abstract}
Conformal theories in a $d$ dimensional spacetime may be expressed as manifestly conformal theories in a $d+2$ dimensional conformal space as first proposed by Dirac. The reduction to $d$ dimensions goes via the $d+1$ dimensional hypercone in the conformal space. Here we give a rather extensive expos\'e of such theories. We review and extend the theory of spinning conformal particles. We give a precise and geometrical formulation of manifestly conformal fields for which we give a consistent action principle. The requirement of invariance under special gauge transformations off the hypercone plays a fundamental role here. Maxwell's theory and linear conformal gravity are derived in the conformal space and are  treated in detail. Finally, we propose  a consistent coordinate invariant action principle  in the conformal space and give an action that should correspond to  conformal gravity.
\end{abstract}\end{center}\end{titlepage}

\setcounter{page}{1}
\setcounter{equation}{0}
\section{Introduction}
In this work we develop the classical properties of manifestly conformally covariant theories including both particle and field theories.

In dimensions $d>2$ the conformal group is $SO(d,2)$ which, however, acts nonlinearly on the Minkowski coordinates. In 1936 Dirac \cite{Dirac:1936}  proposed a manifestly conformally covariant formulation in which the Minkowski coordinates are replaced by coordinates on which $SO(d,2)$ acts linearly. The theory lives then on a $d+1$ dimensional hypercone in a $d+2$ dimensional conformal space. Explicitly he treated scalar, spinor and vector fields in $d=4$. Since then there have appeared many papers on the subject. Among the early papers (see \eg \cite{Mack:1969,Kastrup:1966,Budini:1979}) 
the paper by Mack and Salam \cite{Mack:1969} is  particularly elucidating. It clarifies the projections to Minkowski space in general terms for arbitrary spins in $d=4$. An Euclidean version for  quantum field theory was treated in \eg \cite{Adler:1972,Mack:1973,Ansourian:1977sx}.

Particle models in terms of the manifestly conformal coordinates were proposed in \cite{Marnelius:1979conformal}. Quantization \cite{Marnelius:1980} yields  Dirac's wave equations  in \cite{Dirac:1936}. By means of the result of Bracken and Jessup \cite{Bracken:1982ny} particle models for arbitrary spins were possible to derive (see \cite{Siegel:1988ru,Siegel:1988gd,Siegel:1994cc} and \cite{Marnelius:1990de,Martensson:1992ax}). The particle models are helpful to understand the field theories and have been used in many later papers. Some recent papers on the whole subject  are given in \cite{Codirla:1997,Britto-Pacumio:1999ax,Bars:2006ir,Bars:2006dy,Bars:2000cv,Bars:2001um,Arvidsson:2006}. 

 To begin with  we shall rather extensively  treat and further extend the particle models in \cite{Marnelius:1990de,Martensson:1992ax}.  What concerns particles in external fields, our results are consistent with the corresponding field theories and agree also  with the previous results in \cite{Britto-Pacumio:1999ax,Bars:2000cv,Bars:2001um}.

The emphasis is, however, on manifestly conformal field theories which are considerably developed here. 
We give a precise specification of the conditions to be satisfied by such fields. The properties of the wave functions and the external fields in the conformal particle models serve as a guiding principle for these conditions. However, the field theory is specified in a more geometrical sense in the $d+2$ dimensional conformal space. The restriction to the hypercone is \eg replaced by a requirement of an invariance under special gauge transformations off the hypercone, an invariance which at the end will allow for the restriction to the hypercone as a gauge choice. (These gauge transformations were introduced by Dirac \cite{Dirac:1936}). With the proposed conditions we are able to give a precise and consistent action principle for manifestly conformal fields. Actions for such fields have been discussed before. The first treatment seems to be in  \cite{Budini:1979}. However, there the measure was directly projected to four (easily generalized to $d$) dimensions which prohibits the derivation of manifestly conformally covariant equations, at least naively. In \cite{Marnelius:1980} the invariant measure on the five ($d+1$) dimensional hypercone was proposed for the actions. (Recently, similar actions are also considered in \cite{Bars:2006dy}.) Here we develop this formulation and show that it is consistent provided our conditions on the fields are satisfied and provided we impose special gauge invariance off the hypercone. This gauge invariance is also crucial for the reduction to four ($d$) dimensions of both the actions and the equations of motion. We perform a rather extensive treatment of Maxwell's theory and linear conformal gravity constructed from scratch in the conformal space. (In \cite{Binegar:1982ku} a different approach to linear gravity is given.)  For these models we give consistent actions and equations of motions where the latter follow from the first. We also show that when they are reduced to four dimensional spacetime we get precisely standard Maxwell's theory and linear conformal gravity. 

Our main objective, however, is not just  to clarify the properties of spin two like theories  on the hypercone  but rather to find  a coordinate invariant formulation of conformal gravity in a manifest form in two dimensions higher. This we also accomplish and it is perhaps the main result here. (A previous different proposal for gravity was given in \cite{Preitschopf:1998}.) Actually this work started and  was planned in connection with the Master thesis \cite{Arvidsson:2001} in which manifestly conformally covariant field theories were treated,  particularly was linear conformal gravity studied. Due to health problems of one of us (RM) we could resume our work only this year.

The paper is organized as follows: In section 2 we give the basics of the manifestly conformally covariant formulation in the conformal space and how it is connected to the reduced spacetime. In section 3 we review the conformal particle models in \cite{Marnelius:1979conformal} here given in a  slightly simpler form and generalize them to arbitrary dimensions. In section 4 we treat  the conformal particle in external symmetric tensor fields in arbitrary dimensions. (Previously the vector field was treated in \cite{Marnelius:1979conformal} and the rank two tensor field in \cite{Britto-Pacumio:1999ax,Bars:2000cv}, and arbitrary tensor fields in a different setting in \cite{Bars:2001um}.) In section 5 we give an improved treatment of the conformal particle models  for arbitrary spins given  in \cite{Marnelius:1990de,Martensson:1992ax} generalized to arbitrary dimensions. In section 6 we quantize these models and obtain  free tensor  fields which have not been given before. However, we find problems with the rank two formulation. In section 7 we show that these problems may be understood from the connection between homogeneity and the order of the field equation. In fact, this connection tells us what order of the field equation we should have for symmetric tensor fields of rank two and higher. In section 8 we 
specify the basic ingredients of a more geometrical field theory in the $d+2$ dimensional conformal space. We introduce the special gauge transformations off the hypercone and other defining properties.
 In section 9 we then present our precise action principle  which requires that the actions are invariant under the special gauge transformations. We give its implications for scalar, spinor, and generalized spin one. For symmetric second rank tensor fields in $d=4$ we find that only linearized conformal gravity is allowed by the special gauge invariance. In section 10 we outline what is needed in order to have a coordinate invariant formulation. In section 11 we give the conformal particle in a curved background in external scalar and vector fields  in a coordinate invariant form. (This is a slight generalization of \cite{Britto-Pacumio:1999ax,Bars:2000cv}.) We solve then the derived conditions and the solutions are  completely in agreement with the natural proposals in section 10. In section 12  we give then the generalized action principle for coordinate invariant actions in a curved dynamical background on a generalized hypercone. Then we look for a  possible gravity theory and demonstrate that in $d=4$ the invariance under  generalized special gauge transformations only allow for conformal gravity. Finally we conclude the paper in section 13. In a couple of appendices we give some details for the reductions of the manifestly conformally covariant formulation to $d$ dimensional relativistic theories. In appendix A we give some general aspects of the reduction of the manifestly conformal fields in $d+2$ dimensions to $d$ dimensions. In appendix B we give some details for the reduction of the spin one theory in $d=4$, and in appendix C the corresponding details for linear manifestly covariant gravity in $d=4$. In appendix D we summerize some properties of relativistic particle models for arbitrary spins in $d=4$ generalized to arbitrary dimensions $d$ whose quantum wave functions we expect to correspond to the wave functions from the conformal particle models in section 6.  

\setcounter{equation}{0}
\section{Manifestly conformally covariant formulation}
A large class of massless theories are conformally invariant. In $d>2$ we have apart from Poincar\'e invariance also invariance under scaling, $x^{\mu}\ra \la x^{\mu}$, and under the special conformal transformations
\be
&&x^{\mu}\;\;\ra\;\;x^{\prime\mu}=\frac{x^{\mu}+b^{\mu}x^2}{1+2b\cdot x+b^2x^2}.
\e{01}
All these transformations form the conformal group. Although the transformations are nonlinear the group is simply $SO(d,2)$, a pseudoorthogonal Lie group similar to the Lorentz group $SO(d-1,1)$.
This led Dirac \cite{Dirac:1936} to propose a manifestly conformally covariant formulation for conformal theories. He did this explicitly for free scalar, spinor, and vector fields. This formulation is given on a higher dimensional space called the {\em conformal space}. The coordinates on this $d+2$ dimensional space are denoted $y^A=(y^\mu,y^{d+1},y^{d+2})$, and involve  two time-like directions ($y^0$, $y^{d+2}$).  The indices are raised and lowered by the diagonal metric
\be
&&\eta_{AB}={\rm diag}(1,-1,\ldots,-1,1).
\e{02}
(The opposite sign would be better, but we follow some old literature here. The reductions in appendices B and C are, however, performed using the opposite sign.) $SO(d,2)$ acts linearly on the coordinates $y^A$ and leave the scalar products invariant.  Following Dirac the ordinary spacetime theory is required to live on a $d+1$-dimensional hypercone in the conformal space.
This hypercone is defined through the relation
\be
&&y^2 \equiv \eta_{AB} y^A y^B = 0,
\e{03}
which by definition then is invariant under SO($d$,2) transformations. Due to the projectiveness of this relation the dimensions of $y^A$ may be chosen freely. Choosing dimension length the Minkowski coordinates $x^{\mu}$ are reached by the nonlinear point transformation
\be
x^{\mu} & = & \frac{y^{\mu}}{y^-}R, \qquad y^-\equiv y^{d+2}-y^{d+1}, \nn\\
x^{d+1} & = & y^-,  \nn\\
x^{d+2} & = & \frac{y^2}{R}, 
\e{04}
where $R$ is a constant with dimension length. This transformation is invertible and the inverse is
\be
y^{\mu} & = & \frac{x^{d+1}}{R}x^{\mu}, \nn \\
y^-&=&x^{d+1},\nn\\
y^+&=&\frac{x^{d+2}}{2x^{d+1}}R-\frac{x^{d+1}}{2R}\eta_{\mu\nu}x^{\mu}x^{\nu},\qquad y^+\equiv \half\bigl(y^{d+2}+y^{d+1}\bigr).
\e{05}
This transformation is well defined for $y^-\neq 0$ ($x^{d+1}\neq 0$), which is a topological restriction (see also appendix A). The flat metric \rl{02} induces now an invertible metric $g_{AB}(x)$ in the $x^A$-coordinates whose nonzero elements are
\be
g_{\mu \nu}(x) & = & \ga^2 \eta_{\mu \nu}, \quad \ga\equiv  \frac{x^{d+1}}{R} ,
\nn\\
g_{(d+1)(d+1)}(x) & = & - \frac{x^{d+2}}{R\ga^2},\nn\\
g_{(d+1)(d+2)}(x) & = & \frac{1}{2\ga} = g_{(d+2)(d+1)}(x).
\e{06}

On the hypercone $y^2=0$ ($x^{d+2}=0$) the coordinate of the extra dimension $x^{d+1}$ or $\ga$ in \rl{06} acts as a projective parameter. For the metric above we find a reduction to the Minkowski metric for $x^{d+1}=\pm R$. However, as \eg the particle models show, $\ga=x^{d+1}/R$ may be chosen to be an arbitrary function of $x^{\mu}$ in which case the projected space is turned into an arbitrary conformally flat space. Conformal invariance in the sense of invariance under 
\be
&&g_{\mu\nu}\;\ra\;\la(x) g_{\mu\nu}
\e{07}
is obviously automatic for all conformal theories which are derivable from the manifestly conformally covariant formulation.

How the manifestly conformally covariant fields are reduced to fields on Minkowski space or the conformally flat space is given in appendices A-C.

 \setcounter{equation}{0}
\section{Conformal particle}
In \cite{Marnelius:1979conformal} one of us gave a Lagrangian for a manifestly conformal invariant free particle in $d=4$, both for spin zero, spin one-half and partly spin one. The free spin zero particle may alternatively be described by the slightly simpler Lagrangian
\be
&&L(\tau)={1\over 2v}\dot{y}^2+\la y^2,
\e{201}
where we now consider arbitrary spacetime dimensions $d$. The coordinate $y^A$ is an $SO(d,2)$-vector. Thus, the index $A$ runs here over $d+2$ different values. $v$ is an einbein variable and $\la$ a Lagrange multiplier. The action
\rl{201} is reparametrization invariant and is of the form of an ordinary free massless particle apart from a term that forces the particle to be on the hypercone $y^2=0$. 
%(A gauge fixed form with $v=1$ was considered by Siegel in \cite{Siegel:1988ru,Siegel:1988gd}.)
 A Hamiltonian analysis of \rl{201} yields the first class constraints
\be
&&\chi_1\equiv p^2=0,\quad\chi_2\equiv p\cdot y=0,\quad\chi_3\equiv y^2=0,
\e{202}
together with the trivial primary constraints $p_v=0$ and $p_\la=0$. These constraints satisfy a closed Poisson algebra which is an SL(2,R)-algebra. We have ($\{y^A, p_B\}=\del^A_B$)
\be
&&\{\chi_1, \chi_2\}=-2\chi_1,\quad\{\chi_2, \chi_3\}=-2\chi_3,\quad\{\chi_1, \chi_3\}=-4\chi_2.
\e{2021}
A Dirac quantization of these constraints yields the wave equations (given in \cite{Marnelius:1980} for $d=4$)
\be
&&y^2\Phi(y)=0,\quad\Box\Phi(y)=0,\quad(y\cdot\dif+(d+2)/2)\Phi(y)=0.
\e{203}
The first equation is solved by the ansatz
\be
&&\Phi(y)=\del(y^2)\phi(y).
\e{204}
The remaining equations in \rl{203} require
 then
\be
&&\Box\phi(y)=0,\quad (y\cdot \dif+d/2-1)\phi(y)=0
\e{205}
on the hypercone $y^2=0$. For $d=4$ these equations are identical to the wave equations for a scalar field given by Dirac in \cite{Dirac:1936}.

\subsection{The spinor model}
A corresponding supersymmetric version of the Lagrangian \rl{201} is (cf \cite{Marnelius:1979conformal,Martensson:1992ax})
\be
&&L(\tau)={1\over 2v}\dot{y}^2+{i\over 2}\psi\cdot\dot{\psi}+i\rho y\cdot\psi+\la y^2,
\e{206}
where $\psi^A$ is a real, odd Grassmann variable which also is an $SO(d,2)$-vector. $\rho$ is a real, odd Grassmann variable which acts as an additional Lagrange multiplier to $\la$. A Hamiltonian analysis requires here  $\psi^A$ to satisfy the symmetric Poisson relations ($\eta^{AB}$ is the $SO(d,2)$-metric)
\be
&&\{\psi^A, \psi^B\}=i\eta^{AB}.
\e{207}
The Hamiltonian analysis of \rl{206} yields furthermore the first class constraints
\be
&&\chi_1\equiv p^2=0,\quad\chi_2\equiv p\cdot y=0,\quad\chi_3\equiv y^2=0,\nn\\
&&\chi_4\equiv p\cdot\psi=0,\quad\chi_5\equiv y\cdot\psi=0,
\e{208}
apart from the primary constraints $p_v=0$, $p_\la=0$, and $p_\rho=0$.
Their Poisson algebra is a supersymmetric SL(2,R): They satisfy \rl{2021} and
\be
&&\{\chi_4, \chi_4\}=i\chi_1,\quad\{\chi_5, \chi_5\}=i\chi_3,\quad\{\chi_4, \chi_5\}=i\chi_2,\nn\\
&&\{\chi_4, \chi_1\}=0,\quad\{\chi_4, \chi_2\}=-\chi_4,\quad\{\chi_4, \chi_3\}=-2\chi_5,\nn\\
&&\{\chi_5, \chi_1\}=2\chi_4,\quad\{\chi_5, \chi_2\}=\chi_5,\quad\{\chi_5, \chi_3\}=0.
\e{2081}
A Dirac quantization yields here the wave equations (given in \cite{Marnelius:1980} for $d=4$)
\be
&&\Gamma\cdot\dif\Psi(y)\equiv\dif\!\!\!\slash\Psi(y)=0,\quad y\cdot\Gamma\psi(y)\equiv y\!\!\!\slash\Psi(y)=0,
\e{209}
where $\Gamma^A$ are $\ga$-matrices satisfying the anticommutation relations
\be
&&[\Gamma^A, \Gamma^B]_+=2\eta^{AB}.
\e{210}
They are therefore matrices of dimensions $2^{(d+2)/2}\times 2^{(d+2)/2}$. (Here as well as in the following it is clear that a manifestly conformally invariant formulation only exists for even spacetime dimensions $d$.)
Notice that the equations \rl{209} imply
\be
&&y^2\Psi(y)=0,\quad\Box\Psi(y)=0,\quad(y\cdot\dif+(d+2)/2)\Psi(y)=0,
\e{211}
since $(y\!\!\!\slash)^2=y^2$, $(\dif\!\!\!\slash)^2=\Box$, and $[y\!\!\!\slash, \dif\!\!\!\slash]_+=2y\cdot\dif+d+2$. The second equation in \rl{209} is solved by the ansatz
\be
&&\Psi(y)=\del(y^2)y\!\!\!\slash\psi(y),
\e{212} 
The equations \rl{209} and \rl{211} require then
\be
&&y\!\!\!\slash\dif\!\!\!\slash\psi(y)=0,\quad y\!\!\!\slash(y\cdot\dif+d/2)\psi(y)=0,
\e{213}
on the hypercone $y^2=0$. Later we shall impose the natural  stronger condition
\be
&& (y\cdot\dif+d/2)\psi(y)=0.
\e{214}
For $d=4$ the equations \rl{213} and \rl{214}  agree with Dirac's wave equations in  \cite{Dirac:1936} apart from that Dirac considered a four-spinor instead of an eight-spinor (interpreted as two four-spinors in \cite{Marnelius:1980}). It is always possible to project out a four-spinor by a conformal chiral condition. (For discussions of these properties, see also \cite{Budini:1979} and references therein.)

 \setcounter{equation}{0}
\section{Conformal particle in external fields}
Here we shall consider the spinless particle in the previous section in various external fields.

\subsection{Conformal particle in an external scalar field}
The spinless particle in an external scalar field may be described by the reparametrization invariant action
\be
&&L(\tau)={1\over 2v}\dot{y}^2-\half v g^k\phi^k(y)+\la y^2,
\e{301}
where $\phi$ is the external scalar field, and $g$ a real coupling constant. $k$ is a constant (a positive integer). A Dirac analysis yields here the constraints
\be
&&\chi_1\equiv p^2+g^k\phi^k(y)=0,\quad\chi_2\equiv p\cdot y=0,\quad\chi_3\equiv y^2=0.
\e{302}
%(From the form of $\chi_1$ it follows that the physical Hamiltonian is positive for even $k$.) 
In order for these constraints to satisfy the same algebra \rl{2021} as the free constraints \rl{202} the external field must satisfy the homogeneity condition
\be
&&(y\cdot \dif+{2/ k})\phi(y)=0.
\e{303}
Comparison with the last property in \rl{205} of the free spinless particle we find 
\be
&&k={4\over d-2},
\e{304}
which means that $k=2$ for $d=4$, and $k=1$ for $d=6$, and that $k$ is fractional for $d>6$.

\subsection{Conformal particle in an external vector field}
The spinless particle in an external vector field is described by the reparametrization invariant action \cite{Marnelius:1979conformal}
 \be
&&L(\tau)={1\over 2v}\dot{y}^2- gA_B(y)\dot{y}^B+\la y^2.
\e{305}
A Dirac analysis yields here the constraints:
\be
&&\chi_1\equiv (p+gA(y))^2=0,\quad\chi_2\equiv (p+gA(y))\cdot y=0,\nn\\
&&\chi_3\equiv y^2=0.
\e{306}
These constraints satisfy SL(2,R) provided
\be
&&y^AF_{AB}(y)=0,\quad F_{AB}(y)\equiv\dif_AA_B(y)-\dif_BA_A(y).
\e{307}
This condition is satisfied if we impose \cite{Dirac:1936,Marnelius:1979conformal}
\be
&&y^AA_A(y)=0,\quad(y\cdot\dif+1)A_B(y)=0.
\e{308}
Notice that the second constraint in \rl{306} then becomes $\chi_2\equiv p\cdot y=0$. (The conditions \rl{307} and \rl{308} were also considered by Dirac \cite{Dirac:1936}.)

\subsection{Conformal particle in an external second order tensor field}
The spinless particle in an external second order symmetric tensor field may be described by the reparametrization invariant action
 \be
&&L(\tau)={1\over 2v}\dot{y}^2- {g\over 2v}H_{AB}(y)\dot{y}^A\dot{y}^B+\la y^2,
\e{309}
which may be rewritten as 
 \be
&&L(\tau)={1\over 2v}\dot{y}^AG_{AB}(y)\dot{y}^B+\la y^2,
\e{310}
where
\be
&&G_{AB}(y)\equiv\eta_{AB}-gH_{AB}(y).
\e{311}
We find now the constraints
\be
&&\chi_1\equiv  p_AG^{AB}(y)p_B=0,\quad\chi_2\equiv p_AG^{AB}(y)y_B=0,\nn\\
&&\chi_3\equiv y^2=0,
\e{312}
where $G^{AB}(y)$ and $G_{AB}(y)$ are required to satisfy
\be
&&G^{AB}(y)G_{BC}(y)=\del^A_C.
\e{313}
The condition that these constraints satisfy an SL(2,R)-algebra is then satisfied if ($y^A\equiv\eta^{AB}y_B$)
\be
&&y^A=G^{AB}(y)y_B,\quad y\cdot\dif G^{AB}(y)=0.
\e{314}
For the external field $H_{AB}(y)$ this demands
\be
&&y^AH_{AB}(y)=0,\quad y\cdot\dif H_{AB}(y)=0.
\e{315}

 \subsection{Conformal particle in an external  tensor field of arbitrary order}
The spinless particle in an external  symmetric tensor field of  order  $s$ may be described by the reparametrization invariant action
 \be
&&L(\tau)={1\over 2v}\dot{y}^2- {g\over s!v^{s-1}}H_{A_1\cdots A_s}(y)\dot{y}^{A_1}\cdots\dot{y}^{A_s}+\la y^2.
\e{316}
The conjugate momentum is here
 \be
&&p_A={\dif L(\tau)\over \dif \dot{y}^A}={1\over v}\dot{y}_A-{g\over (s-1)!v^{s-1}}H_{AB_1\cdots B_{s-1}}(y)\dot{y}^{B_1}\cdots\dot{y}^{B_{s-1}}.
\e{317}
 $\dot{y}$ expressed in terms of $p$ may be found by the ansatz
\be
&&\dot{y}^A= v\biggl(p^A+\sum_{n=1}^\infty\Lambda^{AB_1\cdots B_{1+n(s-2)}}(y)p_{B_1}\cdots p_{B_{1+n(s-2)}}\biggr).
\e{318}
This inserted into \rl{317} determines the $\Lambda$'s. We find to the lowest orders
\be
&&\Lambda^{AB_1\cdots B_{s-1}}(y)={g\over(s-1)!}H^{AB_1\cdots B_{s-1}}(y),\nn\\&&
\Lambda^{AB_1\cdots B_{2s-3}}(y)={g^2\over(s-1)!(s-2)!}\biggl(H^{AB_1\cdots B_{s-2}C}(y)H_C^{\;\;\;B_{s-1}\cdots B_{2s-2}}(y)\biggr)_{{\rm sym} B_i},\nn\\
\e{319}
where ${\rm sym} B_i$ means symmetrization in the indices $B_i$, $i=1,\ldots,2s-2$.
It is clear that the $n$th order $\Lambda$ has the form $\Lambda_n=k_{ns}g^nH^n$, where $k_{ns}$ is a constant dependent on $n$ and $s$.
The Lagrangian \rl{316} yields the constraints (we suppress indices)
\be
&&\chi_1\equiv p^2+\sum_{n=1}^\infty f_n(s) g^n H^n(y) p^{ns},\nn\\&&\chi_2\equiv \biggl(p+\half s \sum_{n=1}^\infty nf_n(s)g^n H^n(y) p^{ns-1}\biggr)\cdot y,\quad \chi_3\equiv y^2.
\e{320}
where $f_n(s)$ are constant factors which to lowest orders are
\be
&&f_1(s)={1\over s!},\quad f_2(s)={1\over ((s-1)!)^2}.
\e{321}
The constraints \rl{320} satisfy an SL(2,R)-algebra if
\be
&&H_{A_1\cdots A_{s-1}C}(y)y^C=0,\quad (y\cdot\dif -s+2)H_{A_1\cdots A_{s}}(y)=0,
\e{322}
 independent of the form of the constants $f_n(s)$.
Notice that the first equality implies 
\be
&&\chi_2\equiv p\cdot y
\e{323}
in \rl{320}.
However, notice also that $\chi_1$ for $s>2$ still contains higher order powers than two in $p$.  This means that the equations of the wave functions will be of higher order than two for $s>2$. The result \rl{322}, which also was obtained from a different setting in \cite{Bars:2001um}, agrees with the previous results for $s=1$ and $s=2$.

 \setcounter{equation}{0}
\section{Conformal particles of arbitrary spins generalized to arbitrary dimensions}
Manifestly conformal particle models in $d=4$ for arbitrary spins were constructed in \cite{Marnelius:1990de,Martensson:1992ax}. These models are analogous to the relativistic particle models of arbitrary spins given in \cite{Marnelius:1988ab} (cf.\cite{Gershun:1979fb,Howe:1988ft}) partly reviewed in appendix D. Their Lagrangians are $O(N)$ extended supersymmetric models which describe a free particle with spin $s=N/2$. The geometrical Lagrangian for these manifestly conformal particles is given in \cite{Martensson:1992ax}, which  also could be simplified along the line of section 3. The Lagrangian \rl{206} is then the $s=1/2$ model. In the following we generalize the models for arbitrary $s$ to arbitrary spacetime dimensions $d$, and refer to them as the $s=N/2$ theory when arbitrary $d$ is considered. Notice, however, that they describe a spin $s$ particle only in $d=4$. The scalar  model in section 3 may also be called the $s=0$ model.

The derivation of the conformal models in \cite{Marnelius:1990de,Martensson:1992ax} were performed using the results of Bracken and Jessup in \cite{Bracken:1982ny}. Here we follow this derivation for arbitrary $d$ and derive the Dirac quantization of these models in a straight-forward manner. 
(Previously a BRST quantization was performed in \cite{Martensson:1992ax}.)

 Bracken and Jessup had shown that any model satisfying the massless Klein-Gordon equation is manifestly conformally invariant if
(all indices are $SO(d,2)$-indices.)
\be
&&W_{AB}|\psi\hb=0, \quad W_{AB}^{\dagger}=W_{AB},
\e{501}
where 
\be
&&W_{AB}\equiv J_{AC}J^C_{\;\;\;B}+J_{BC}J^C_{\;\;\;A}+{2\over d+2}\eta_{AB}J_{CD}J^{CD}=W_{BA},\quad W^A_{\;\;\;A}=0,\nn\\
\e{502}
where in turn $J_{AB}$ is the total angular momentum for the particle in $d+2$ dimensions.  We have
\be
&&J_{AB}=L_{AB}+S_{AB}=-J_{BA}, \quad L_{AB}=y_Ap_B-y_Bp_A,
\e{503}
where $S_{AB}$ represents the spin degrees of freedom. Notice that $y^A$, $p_A$, and $S_{AB}$ are operators satisfying
\be
&&[y^A, p_B]=i\del^A_B,\quad [S_{AB}, y^C]=[S_{AB}, p_C]=0.
\e{504}
 All this was originally formulated for $d=4$. Here we  formally extend these  results to arbitrary dimensions $d$. $J_{AB}$, $L_{AB}$ and $S_{AB}$ satisfy the $SO(d,2)$-algebra:
\be
&&[J_{AB}, J_{CD}]=i(\eta_{AD}J_{BC}+\eta_{BC}J_{AD}-\eta_{AC}J_{BD}-\eta_{BD}J_{AC}).
\e{5041}
If we define $W_{AB}^{(J)}\equiv W_{AB}$, then we may write
\be
&&W_{AB}=W_{AB}^{(L)}+W_{AB}^{(L,S)}+W_{AB}^{(S)},
\e{505}
where $W_{AB}^{(L)}$ and $W_{AB}^{(S)}$ are analogous to $W_{AB}^{(J)}$, and where
\be
&&W_{AB}^{(L,S)}\equiv 2\biggl(L_{AC}S^C_{\;\;\;B}+L_{BC}S^C_{\;\;\;A}-{2\over d+2}\eta_{AB} L_{CD}S^{DC}\biggr).
\e{506}
The different parts in \rl{505} are of different degrees in $S_{AB}$. It is therefore natural to look for a solution of \rl{501} which also satisfies
\be
&&W_{AB}^{(L)}|\psi\hb=0,\quad W_{AB}^{(L,S)}|\psi\hb=0,\quad W_{AB}^{(S)}|\psi\hb=0.
\e{507}
We may then always extract some elementary constraints from these conditions which we may use to define the manifestly conformally invariant particle. We notice first that
\be
&&W_{AB}^{(L)}=(y_Ap_B+y_Bp_A+p_Ay_B+p_By_A)D-2y_Ay_Bp^2-2p_Ap_By^2+\nn\\
&&+{2\over d+2}\eta_{AB}(y^2p^2+p^2y^2)-{4\over d+2}\eta_{AB}D^2,
\e{508}
where
\be
&&D\equiv \half (y\cdot p+p\cdot y).
\e{509}
Thus,
\be
&&p^2|\psi\hb=0,\quad y^2|\psi\hb=0,\quad D|\psi\hb=0\quad \Rightarrow\quad W_{AB}^{(L)}|\psi\hb=0.
\e{510}
These conditions are identical to the quantum conditions \rl{203} of the spinless particle as it should be for $S_{AB}=0$. Next we consider some representations of $S_{AB}$. We start with the multi-fermionic formulation
\be
&&S^{AB}\equiv \half i\sum_{j=1}^N(\psi^B_j\psi^A_j-\psi^A_j\psi^B_j),
\e{511}
where the operators $\psi^A_j$ are hermitian (and Grassmann odd)  and satisfy the anticommutation relation
\be
&&[\psi^A_i, \psi^B_j]_+=-\del_{ij}\eta^{AB}.
\e{512}
We find then
\be
&&W_{AB}^{(L,S)}=2i\sum_j(y_A\psi_{jB}+y_B\psi_{jA})(\psi_j\cdot p)-\nn\\&&-2i\sum_j(\psi_{jB} p_A+\psi_{jA} p_B)(\psi_j\cdot y)-\nn\\
&&-{4i\over d+2}\eta_{AB}\sum_j\biggl((y\cdot\psi_j)(\psi_j\cdot p)-(\psi_j\cdot p)(\psi_j\cdot y)\biggr).
\e{513}
Hence,
\be
&&\psi_j\cdot p|\psi\hb=0,\quad \psi_j\cdot y|\psi\hb=0\quad\Rightarrow\quad W_{AB}^{(L,S)}|\psi\hb=0.
\e{514}
Finally, we have
\be
&&W_{AB}^{(S)}=-\sum_{l\neq k}(\psi_{kA}\psi_{lB}+\psi_{kB}\psi_{lA})\psi_{kC}\psi_l^C+{2\over d+2}\eta_{AB}\sum_{l\neq k}(\psi_{kD}\psi_l^D\psi_{kC}\psi_l^C).\nn\\
\e{515}
Hence,
\be
&&(\psi_{kC}\psi_l^C-\psi_{lC}\psi_k^C)|\psi\hb=0\;\;\forall l,k\quad\Rightarrow\quad W_{AB}^{(S)}|\psi\hb=0.
\e{516}
For the multi-fermionic  representation \rl{511} we have then ($\hat{\chi}_n$ are hermitian)
\be
&&\hat{\chi}_n|\psi\hb=0\quad\Rightarrow\quad W_{AB}|\psi\hb=0,
\e{517}
where
\be
&&\hat{\chi}_1\equiv p^2,\quad\hat{\chi}_2\equiv D,\quad \hat{\chi}_3\equiv y^2,
\e{518}
\be
&&\hat{\chi}_{3+j}\equiv \psi_j\cdot p,\quad \hat{\chi}_{3+N+j}\equiv\psi_j\cdot y,
\e{519}
\be
&&\hat{\chi}_{3+2N+[ij]}\equiv \half i(\psi_i\cdot\psi_j-\psi_j\cdot\psi_i).
\e{520}
The last constraint operators are antisymmetric in $i$ and $j$ and  their number is $N(N-1)/2$. The commutator algebra of these $\hat{\chi}_n$ is an $O(N)$-extended supersymmetric  $SL(2,R)$-algebra. The corresponding reparametrization invariant particle model is given by the Hamiltonian
\be
&&H=\sum_n\la_n\chi_n,
\e{521}
where $\chi_n$ are the classical counterparts to the operators \rl{518}-\rl{520}. 
% (The derivation given in \cite{Marnelius:1990de} was performed in a different way.)
 The geometrical Lagrangian and further properties  for $d=4$ are given in \cite{Martensson:1992ax}. 

For $N=1$ we have the spinor model  treated in section 3. In the multi-fermionic case we find instead of \rl{212} $\Psi(y)=\del(y^2)(\prod_i y\!\!\!\slash_i )\psi(y)$, where the multi-spinor $\psi(y)$ then satisfies the homogeneity property $(y\cdot\dif+d/2+1-N)\psi(y)=0$ (cf \cite{Marnelius:1990de}). We expect such a multispinor  to represent spin $s=N/2$ in $d=4$, since it should correspond to the relativistic multispinor model in \cite{Marnelius:1988ab} (see also \cite{Gershun:1979fb, Howe:1988ft}).

One may now derive all possible manifestly conformally invariant spinning particle models by inserting various forms for $S_{AB}$ in the formulas above.

 \setcounter{equation}{0}
\section{Free conformal tensor fields from the conformal particle models}
For $N$ even in \rl{511} we may derive manifestly conformally covariant tensor fields from the general quantization procedure of the preceding section. (We expect these tensor fields to correspond to the similar tensor fields  which follow from  the corresponding quantization of the relativistic particle models reviewed in appendix D.)  What we have to do is simply to introduce
nonhermitian operators defined by
\be
&&b_k^A\equiv{1\over\sqrt{2}}\bigl(\psi^A_{(2k-1)}-i\psi^A_{2k}\bigr),\quad k=1,\ldots,s\equiv N/2,\;\;(s=1,2,\ldots),
\e{601}
which after quantization satisfy the anticommutation relations
\be
&&[b_{iA}, b^{\dagger}_{jB}]_+=-\del_{ij}\eta_{AB},
\e{602}
due to \rl{512}.
In terms of these operators $S_{AB}$ in \rl{511} becomes 
\be
&&S^{AB}=i\sum_{j=1}^s(b^{B\dagger}_jb^A_j-b^{A\dagger}_jb^B_j).
\e{603}
As before we solve the condition \rl{501} by \rl{507}. \rl{510} is still valid.  \rl{514} and \rl{516} have to be rearranged.   In terms of the oscillators \rl{601}, \rl{514} acquires the form  
\be
&& p\cdot b_j|\psi\hb=0,\quad p\cdot b^{\dagger}_j|\psi\hb=0,\quad y\cdot b_j|\psi\hb=0\quad y\cdot b^{\dagger}_j|\psi\hb=0\nn\\
&&\Rightarrow\quad W_{AB}^{(L,S)}|\psi\hb=0.
\e{604}
and \rl{516} becomes
\be
&&(b_{kC}b_l^C-b_{lC}b_k^C)|\psi\hb=0\quad(b^{\dagger}_{kC}b_l^{C\dagger}-b^{\dagger}_{lC}b_k^{C\dagger})|\psi\hb=0\nn\\
&&(b^{\dagger}_{kC}b_l^C-b_{lC}b_k^{C\dagger})|\psi\hb=0\;\;\forall l,k\quad\Rightarrow\quad W_{AB}^{(S)}|\psi\hb=0.
\e{605}
The algebra of the elementary constraints is of course identical to what we had in the previous section.

We may solve the conditions \rl{510}, \rl{604} and \rl{605} by means of the following general ansatz of the state vector: (the sum is over all possible values of $n_j$, $j=1,\ldots,s$)
\be
&&|\psi\hb\equiv|\cF\hb=\sum_{n_j}\cF_{A_1\cdots A_{n_1};B_1\cdots B_{n_2}; C_1\cdots C_{n_3}; \cdots    }(y)|0\hb^{A_1\cdots A_{n_1};B_1\cdots B_{n_2};C_1\cdots C_{n_3};\cdots},\nn\\
\e{606}
where
\be
&&|0\hb^{A_1\cdots A_{n_1};B_1\cdots B_{n_2};C_1\cdots C_{n_3};\cdots}
\equiv b_1^{A_1\dagger}\cdots b_1^{A_{n_1}\dagger}b_2^{B_1\dagger}\cdots b_2^{B_{n_2}\dagger}b_3^{C_1\dagger}\cdots b_3^{C_{n_3}\dagger}\cdots |0\hb,\nn\\
&& b_j^A|0\hb=0,\quad p_A|0\hb=0.
\e{607}
The last conditions in \rl{605} contain the restrictions
\be
&&(b^{\dagger}_{jC}b_j^C+(d+2)/2)|\cF\hb=0,
\e{608}
which implies that the $s$ field contains $s$ sets of $(d+2)/2$ antisymmetric indices (\ie $n_j=(d+2)/2$ in \rl{606}). From \rl{605} we have furthermore
\be
&&(b_{kC}b_l^C-b_{lC}b_k^C)|\psi\hb=0\quad(b^{\dagger}_{kC}b_l^{C\dagger}-b^{\dagger}_{lC}b_k^{C\dagger})|\psi\hb=0\nn\\
&&(b^{\dagger}_{kC}b_l^C-b_{lC}b_k^{C\dagger})|\psi\hb=0,\quad\forall l\neq k,
\e{609}
which yields the following symmetry properties of the tensor field: $\cF_{\cdots}$ is symmetric under interchange of any two groups of $(d+2)/2$ antisymmetric indices. Furthermore, all contractions over the groups are zero: ($n=(d+2)/2$ in the following)
\be
&&\cF^{\quad\quad\quad\;\;A}_{AA_2\cdots A_{n};\;\;\;B_2\cdots B_{n};C_1\cdots C_{n};\cdots}(y)=0,\quad {\rm etc.}
\e{610}
From \rl{510} we find
\be
&&y^2|\cF\hb=0\quad\Leftrightarrow\quad y^2\cF_{A_1\cdots A_{n};B_1\cdots B_n;C_1\cdots C_n;\cdots}(y)=0,\nn\\
&&p^2|\cF\hb=0\quad\Leftrightarrow\quad \Box \cF_{A_1\cdots A_{n};B_1\cdots B_n;C_1\cdots C_n;\cdots}(y)=0,\nn\\
&&D|\cF\hb=0\quad\Leftrightarrow\quad (y\cdot\dif+(d+2)/2)\cF_{A_1\cdots A_{n};B_1\cdots B_n;C_1\cdots C_n;\cdots}(y)=0.\nn\\
\e{611}

 Finally, \rl{604} yields
\be
&&p\cdot b_j|\cF\hb=0\quad\Leftrightarrow\quad \dif^A\cF_{AA_2\cdots A_n;B_1\cdots B_n;C_1\cdots C_n;\cdots}(y)=0,\; {\rm etc},
\e{614a}
\be
&&p\cdot b_j^{\dagger}|\cF\hb=0\quad\Leftrightarrow\quad \dif_{[A_1}\cF_{A_2\cdots A_{n+1}];B_1\cdots B_n;C_1\cdots C_n;\cdots}(y)=0,\; {\rm etc},
\e{614b}
\be
&&y\cdot b_j|\cF\hb=0\quad\Leftrightarrow\quad y^A\cF_{A A_2\cdots A_n;B_1\cdots B_n;C_1\cdots C_n;\cdots}(y)=0,\; {\rm etc},
\e{614c}
\be
&&y\cdot b_j^{\dagger}|\cF\hb=0\quad\Leftrightarrow\quad y_{[A_1}\cF_{A_2\cdots A_{n+1}];B_1\cdots B_n;C_1\cdots C_n;\cdots}(y)=0,\; {\rm etc},
\e{614d}
where $n\equiv (d+2)/2$, and where etc means the same relations for the $B$-indices, $C$-indices etc.

The first condition in \rl{611} is solved by
\be
&&\cF_{A_1\cdots A_{n};B_1\cdots B_n;C_1\cdots C_n;\cdots}(y)=\del(y^2)F_{A_1\cdots A_{n};B_1\cdots B_n;C_1\cdots C_n;\cdots}(y),
\e{612}
and the remaining two conditions require then
\be
&&\Box F_{A_1\cdots A_{n};B_1\cdots B_n;C_1\cdots C_n;\cdots}(y)=0,
\e{6131}
\be
&&(y\cdot\dif+d/2-1)F_{A_1\cdots A_{n};B_1\cdots B_n;C_1\cdots C_n;\cdots}(y)=0,
\e{6132}
on the hypercone $y^2=0$. Notice that $F_{\cdots}$ in \rl{612} is ambiguously defined since \rl{612} is invariant under the transformation
\be
&&F_{A_1\cdots A_{n};B_1\cdots B_n;C_1\cdots C_n;\cdots}(y)\quad\rightarrow\nn\\
&&F_{A_1\cdots A_{n};B_1\cdots B_n;C_1\cdots C_n;\cdots}(y)+y^2\tilde{F}_{A_1\cdots A_{n};B_1\cdots B_n;C_1\cdots C_n;\cdots}(y),
\e{6133}
for arbitrary functions $\tilde{F}_{\cdots}$. This ambiguity will be discussed in section 8. 

Below we give some simple solutions of these equations:

\subsection{$s=1$ theories}
In this case we have only one anticommuting oscillator $b^A$, which means that we have only one set of  $(d+2)/2$ antisymmetric indices.  Below we treat $d=2,4,6$,  and give the structure for arbitrary dimensions $d$.

\subsubsection{$s=1$ in $d=2$}
The basic field is here $\cF_{AB}$ due to \rl{608}. The conditions \rl{611} imply
\be
&&\cF_{AB}(y)=\del(y^2)F_{AB}(y),
\e{6141}
and 
\be
&& \Box F_{AB}(y)=0,\quad y\cdot\dif F_{AB}(y)=0,
\e{61411}
on the hypercone. In the following we will omit the statements "on the hypercone", since most equations are ambiguous off the hypercone.  (Later in the field theory case in section 8 we will discuss the precise properties of the equations.)
Condition \rl{614d}  is solved by
\be
&&F_{AB}(y)=y_AF_{B}(y)-y_BF_{A}(y),\quad(y\cdot\dif+1)F_A(y)=0,
\e{6142}
and condition \rl{614c} requires
\be
&& y^AF_{AB}(y)=0\quad\Rightarrow\quad  y^AF_{A}(y)=0.
\e{6144}
Condition \rl{614b} is solved by
\be
&&F_{A}(y)=\dif_A\phi(y),\quad y\cdot\dif\phi(y)=0,
\e{6143}
where the last expression follows from the last relation in \rl{6144}.
It is consistent with  the previous homogeneity of scalar fields in $d=2$.
Condition \rl{614a} yields
\be
&&\dif^AF_{AB}(y)=2F_B(y),
\e{6145}
if the last relation in \rl{6144} is valid even off the hypercone, and if
\be
&& \dif^AF_{A}(y)=0\quad\Leftrightarrow\quad \Box\phi(y)=0
\e{6146}
 which is  consistent with the first relation in \rl{61411}.

There are several ambiguities in the above relations. Apart from the general ambiguity off the hypercone the definition of $F_A$ in \rl{6142} is invariant under the transformation
\be
&&F_A(y)\quad\rightarrow\quad F_A(y)+y_AU(y),\quad (y\cdot\dif+2)U(y)=0.
\e{6147}
This ambiguity is partly fixed by the first relation in \rl{6143}. Notice, however, that
the transformation
\be
&&\phi(y)\quad\ra\quad\phi(y)+\half y^2U(y)
\e{61471}
reproduces \rl{6147} on the hypercone. A further ambiguity enters the equations of motion.
The first equation in \rl{61411} requires 
\be
&&\Box F_A(y)=y_AS(y),\quad (y\cdot\dif+4)S(y)=0,
\e{6149}
where $S$ is an arbitrary function. However, this property
 is  obviously connected to the arbitrariness off the hypercone of \rl{6146} from \rl{6143}.

\subsubsection{$s=1$ in $d=4$ (spin one)}
The basic field is here $\cF_{ABC}$ due to \rl{608}. The conditions \rl{611} imply
\be
&&\cF_{ABC}(y)=\del(y^2)F_{ABC}(y),\quad \Box F_{ABC}(y)=0,\quad (y\cdot\dif+1)F_{ABC}(y)=0.\nn\\
\e{615}
Condition \rl{614d}  yields the solution
\be
&&F_{ABC}(y)=y_AF_{BC}(y)+y_BF_{CA}(y)+y_CF_{AB}(y),\nn\\
&&(y\cdot\dif+2)F_{AB}(y)=0,
\e{616}
and \rl{614b} is then solved by
\be
&&F_{AB}(y)=\dif_AA_B(y)-\dif_BA_A(y).
\e{617}
Eq.\rl{616}
may therefore alternatively be written as
\be
&&F_{ABC}(y)=L_{AB}A_C(y)+L_{BC}A_A(y)+L_{CA}A_B(y),
\e{618}
where
\be
&&L_{AB}\equiv y_A\dif_B-y_B\dif_A.
\e{619}
The  condition \rl{614c}  yields from \rl{616}
\be
&&y^AF_{ABC}(y)=0\quad\Rightarrow\quad y^AF_{AB}(y)=y_BK(y),
\e{620}
which from \rl{617} and the last property in \rl{615} implies
\be
&&y^AA_A(y)=0,\quad(y\cdot\dif+1)A_B(y)=0,
\e{621}
which is in agreement with the external  field result \rl{308}. Notice that $\dif_B(y^AA_A)=-y_BK$.

Apart from the general ambiguity off the hypercone we notice that \rl{616} is invariant under the transformations
\be
&&F_{AB}(y)\quad\ra\quad F'_{AB}(y)=F_{AB}(y)+y_AU_B(y)-y_BU_A(y),\nn\\
&&(y\cdot\dif+3)U_A(y)=0
\e{6211}
for arbitrary functions $U_A$. The expression \rl{617} partly fix this invariance. Notice, however, that the transformation
\be
&&A_B(y)\quad\ra\quad A'_B(y)=A_B(y)-y_B\phi(y)
\e{6212}
yields
\be
&&F_{AB}(y)\quad\ra\quad F'_{AB}(y)=F_{AB}(y)+L_{AB}\phi(y),
\e{6213}
which is of the form \rl{6211} ($U_A=\dif_A\phi$). Thus, \rl{6212} is a gauge transformation in the theory.
That $F_{ABC}$ is left invariant by \rl{6212} may also be seen directly from \rl{618}. We have also the standard gauge invariance under
\be
&&A_B(y)\quad\ra\quad A'_B(y)=A_B(y)+\dif_B\Lambda(y)
\e{6214}
for arbitrary functions $\Lambda$, since \rl{6214} leave $F_{AB}$ invariant.

From \rl{615} and \rl{616} we find
\be
&&\Box F_{ABC}(y)=0\quad \Rightarrow\quad\Box F_{AB}(y)=y_AS_B(y)-y_BS_A(y),
\e{6215}
where $S_A$ are arbitrary functions.
Eq.\rl{614a} yields, if we assume that the first relation in \rl{621} is valid even off the hypercone (a choice to be made in section 8),
\be
&&\dif^AF_{ABC}(y)=2F_{BC}(y)\quad \Rightarrow\quad\dif^AF_{AB}(y)=y_BS(y).
\e{622}
The functions $S_A$ and $S$ reduce the number of equations for $F_{AB}$. They are arbitrary unphysical  functions which only are restricted by the self-consistency of \rl{6215} and \rl{622} which requires
\be
&&(y\cdot\dif+5)S_A(y)=0,\quad (y\cdot\dif+4)S(y)=0.
\e{6221}
In fact, they are related. From $\Box(y^AF_{AB})=0$ again assuming the first relation in \rl{621} to be valid even off the hypercone we find
\be
&&S(y)=\half y^AS_A(y).
\e{6222}
The unphysical  functions $S_A$ and $S$ do not appear in connection to 
the field $F_{ABC}$ in \rl{616} or equivalently \rl{618}. This field is therefore  insensitive to the values of the $S$-functions.  The field $F_{ABC}$ has previously been treated in \cite{Adler:1972,Codirla:1997}.

\subsubsection{$s=1$ in $d=6$}
In $d=6$ we have a field with four antisymmetric indices due to \rl{608}. The conditions \rl{611}
yield
\be
&&\cF_{ABCD}(y)=\del(y^2)F_{ABCD}(y), \quad \Box F_{ABCD}(y)=0, \nn\\&&(y\cdot\dif+2)F_{ABCD}(y)=0.
\e{623}
 Condition  \rl{614d} yields the solution
\be
&&F_{ABCD}(y)=y_AF_{BCD}(y)-y_BF_{CDA}(y)+y_CF_{DAB}(y)-y_DF_{ABC}(y),\nn\\
\e{624}
where $F_{ABC}$ is totally antisymmetric. The homogeneity in \rl{623} requires
\be
&& (y\cdot\dif+3)F_{ABC}(y)=0.
\e{6241}
Condition \rl{614c}  yields 
\be
&&y^AF_{ABCD}(y)=0\quad\Rightarrow\quad y^AF_{ABC}(y)=y_BK_C(y)-y_CK_B(y),\nn\\ &&y^AK_A(y)=0.
\e{625}
Condition \rl{614b} may be written as
\be
&&\dif_AF_{BCD}(y)-\dif_BF_{CDA}(y)+\dif_CF_{DAB}(y)-\dif_DF_{ABC}(y)=0,
\e{626}
and this condition is solved by the expression
\be
&&F_{ABC}(y)=\dif_AA_{BC}(y)+\dif_BA_{CA}(y)+\dif_CA_{AB}(y),
\e{627}
where $A_{BC}$ is antisymmetric. \rl{6241}  and \rl{625} imply then
\be
&&y^AA_{AB}(y)=0,\quad (y\cdot\dif+2)A_{BC}(y)=0,
\e{628}
since 
\be
&&y^AF_{ABC}(y)=-\dif_B(y^AA_{AC}(y))+\dif_C(y^AA_{AB}(y)).
\e{6281}
Eq.\rl{627} inserted into \rl{624} leads to the following alternative expression for the original field
\be
&&F_{ABCD}(y)=L_{AB}A_{CD}(y)+L_{AC}A_{DB}(y)+L_{AD}A_{BC}(y)+\nn\\&&+L_{BC}A_{AD}(y)+L_{BD}A_{CA}(y)+L_{CD}A_{AB}(y),
\e{629}
where $L_{AB}$ is given by \rl{619}.

Apart from the general ambiguity off the hypercone we notice that \rl{624} is invariant under the transformation
\be
&&F_{ABC}(y)\quad\ra\quad F_{ABC}(y)+y_AU_{BC}(y)+y_BU_{CA}(y)+y_CU_{AB}(y)\nn\\
\e{6295}
for arbitrary antisymmetric functions $U_{AB}$. This invariance is partly fixed if we choose $F_{ABC}$ to be of the form \rl{627}. Notice, however, that the transformation
\be
&&A_{AB}(y)\quad\ra\quad A_{AB}(y)-y_AV_B(y)+y_BV_A(y),\quad y^AV_A(y)=0
\e{6292}
yields
\be
&&F_{ABC}(y)\quad\ra\quad F_{ABC}(y)+L_{AB}V_C(y)+L_{BC}V_A(y)+L_{CA}V_B(y),\nn\\
\e{6293}
which is of the form \rl{6295} with $U_{AB}=\dif_AV_B-\dif_BV_A$. Thus, \rl{6292} is a gauge transformation in the theory. One may also easily check that the expression \rl{629} is invariant under \rl{6292}. Apart from this gauge invariance we also have invariance under
\be
&&A_{AB}(y)\quad\ra\quad A_{AB}(y)+\dif_A\Lambda_B(y)-\dif_B\Lambda_A(y)
\e{6294}
for arbitrary functions $\Lambda_A$.

From \rl{624} we find that the equation in \rl{623} requires
\be
&&\Box F_{ABC}(y)=y_AS_{BC}(y)+y_BS_{CA}(y)+y_CS_{AB}(y),
\e{6242}
where $S_{AB}$ are arbitrary antisymmetric  functions. Self-consistency of \rl{6242} requires
\be
&&(y\cdot\dif+6)S_{ABC}(y)=0.
\e{6243}
Assuming the first relation in \rl{628} to be valid even off the hypercone we find also
\be
&&\dif^AF_{ABCD}(y)=2F_{BCD}(y),\nn\\&& \dif^AF_{ABC}(y)=y_BS_C(y)-y_CS_B(y),
\e{6244}
where
\be
&&S_A(y)=\half y^BS_{BA}(y).
\e{6245}

\subsubsection{$s=1$ in arbitrary even dimensions $d$}
In arbitrary even dimensions $d$ the $s=1$ theory leads to  a field with $d/2+1$ antisymmetric indices. Condition \rl{614d} is then solved by expressing this field in terms of fields with $d/2$ antisymmetric indices and the coordinate $y^A$ linearly as in \rl{616} and \rl{624}. Condition \rl{614b} may then be solved by writing this latter fields as follows
\be
&&F_{ABC\cdots}(y)=\sum_{antisym (ABC\cdots)}\dif_AA_{BC\cdots}(y).
\e{6296}
where $A_{ABC\cdots}(y)$ are antisymmetric fields with $d/2-1$ indices as in \rl{617} and \rl{627}. The homogeneity \rl{6132} requires then
\be
&&(y\cdot\dif+d/2)F_{ABC\cdots}(y)=0,\quad (y\cdot\dif+d/2-1)A_{BC\cdots}(y)=0.
\e{6297}
We notice that \rl{6296} yields
\be
&&y^AF_{ABC\cdots}(y)=\sum_{antisym (BC\cdots)}\dif_B(y^AA_{BC\cdots A}(y)).
\e{6298}
Condition \rl{614c} may therefore always be solved by requiring
\be
&&y^AA_{ABC\cdots}(y)=0.
\e{6299}
If this relation is valid even off the hypercone then the field \rl{6296} satisfies the following properties
\be
&&y^AF_{ABC\cdots}(y)=0,\nn\\
&&\Box F_{ABC\cdots}(y)=\sum_{antisym(ABC\cdots)}y_AS_{BC\cdots}(y), \nn\\&&\dif^AF_{ABC\cdots}(y)=\sum_{antisym(BC\cdots)}y_BS_{CD\cdots}(y),\nn\\
&& (y\cdot\dif+d/2+3)S_{BC\cdots}(y)=0,\nn\\
&&(y\cdot\dif+d/2+2)S_{CD\cdots}(y)=0,\nn\\
&&S_{CD\cdots}(y)=\half y^BS_{BCD\cdots}(y),
\e{6291}
where the arbitrary unphysical  functions $S_{ABC\cdots}$ are totally antisymmetric. 
The values of the $S$-functions do not affect the original fields with $d/2+1$ antisymmetric indices.

Further properties of the $s=1$ theory will be discussed in section 8.

\subsection{$s=2$ in $d=4$}
For $s=2$  in $d=4$ which should be a spin two model we have a tensor field with two sets of three antisymmetric indices due to the ansatz \rl{606} and the conditions \rl{608}. We set 
\be
&&\cR_{ABCDEF}(y)\equiv\cF_{ABC;DEF}(y)
\e{630}
in the following. Conditions \rl{611} yield
\be
&&\cR_{ABCDEF}(y)=\del(y^2)R_{ABCDEF}(y),\quad\Box R_{ABCDEF}(y)=0,\nn\\&&(y\cdot\dif+1)R_{ABCDEF}(y)=0.
\e{631}
The conditions \rl{614d} are here solved by the expression
\be
&&R_{ABCDEF}(y)=\sum_{antisym(ABC)}\sum_{antisym(DEF)}y_AR_{BCDE}(y)y_F=   \nn\\
&&y_AR_{BCDE}(y)y_F+y_BR_{CADE}(y)y_F+y_CR_{ABDE}(y)y_F+\nn\\&&
+y_AR_{BCFD}(y)y_E+y_BR_{CAFD}(y)y_E+y_CR_{ABFD}(y)y_E+\nn\\&&
+y_AR_{BCEF}(y)y_D+y_BR_{CAEF}(y)y_D+y_CR_{ABEF}(y)y_D,
\e{632}
where the field $R_{BCDE}$ is antisymmetric in {\sl\sc bc} and in {\sl\sc de}, and satisfies the homogeneity condition
\be
&& (y\cdot\dif+3)R_{ABCD}(y)=0
\e{638}
from \rl{631}.
Two of the conditions in \rl{609} require
\be
&&R_{ABCDEF}(y)-R_{BCDAEF}(y)+R_{CDABEF}(y)-R_{DABCEF}(y)=0,\nn\\
&&R_{ABCDEF}(y)-R_{ABDEFC}(y)+R_{ABEFCD}(y)-R_{ABFCDE}(y)=0,\nn\\
&&\Rightarrow\quad R_{ABCDEF}(y)=R_{DEFABC}(y),
\e{634}
which may be solved by the conditions
\be
&&R_{ABCD}(y)+R_{BCAD}(y)+R_{CABD}(y)=0,\nn\\
&&R_{ABCD}(y)+R_{ACDB}(y)+R_{ADBC}(y)=0,\nn\\
&&\Rightarrow\quad R_{ABCD}(y)=R_{CDAB}(y).
\e{636}
The field $R_{ABCD}$ has then the symmetry properties of a Riemann tensor. Notice, however, that 
$R_{ABCD}$ is not uniquely defined by \rl{632} and \rl{636}. In fact, these expressions are invariant under the following transformation
\be
&&R_{ABCD}(y)\quad\ra\quad R_{ABCD}(y)+y_AU_{BCD}(y)-y_BU_{ACD}(y)+\nn\\
&&+y_CU_{DAB}(y)-y_DU_{CAB}(y),
\e{6321}
where $U_{BCD}$ are arbitrary functions satisfying the properties
\be
&& U_{ABC}(y)=-U_{ACB}(y), \quad (y\cdot\dif+4)U_{ABC}(y)=0,    \nn\\
&&U_{ABC}(y)+U_{BCA}(y)+U_{CAB}(y)=0.
\e{6322}
Condition \rl{614c} requires then 
\be
&&y^AR_{ABCDEF}(y)=0\quad\Rightarrow\quad y^AR_{ABCD}(y)=0,
\e{633}
up to the arbitrariness of the $U$-functions in \rl{6321}.
Conditions \rl{614b} may be solved by the conditions
\be
&&\dif_AR_{BCDE}(y)+\dif_BR_{CADE}(y)+\dif_CR_{ABDE}(y)=0,\nn\\
&&\dif_CR_{ABDE}(y)+\dif_DR_{ABEC}(y)+\dif_ER_{ABCD}(y)=0,
\e{637}
which have the solution
 \be
 &&R_{ABCD}(y)=\half\biggl(\dif_A\dif_C H_{BD}(y)+\nn\\&&+\dif_B\dif_DH_{AC}(y)-\dif_B\dif_CH_{AD}(y)-\dif_A\dif_DH_{BC}(y)\biggr),
 \e{6371}
 which is the linearized Riemann tensor.

The remaining conditions have to do with the equations of motion. The
conditions \rl{609} requires in addition to \rl{636}
\be
&&R^A_{\;\;\:BCDEA}(y)=0,
\e{6372}
which requires
\be
&&R^A_{\;\;\;BCA}(y)=y_BS_C(y)-y_CS_B(y),\nn\\
&&y^AS_A(y)=0,\quad (y\cdot\dif+4)S_A(y)=0.
\e{635}
The arbitrary functions $S_A$ are unphysical  since they do not affect the original tensor \rl{630}. 
The equation of motion in \rl{631} requires
\be
&&\Box R_{ABCD}(y)=y_AS_{BCD}(y)-y_BS_{ACD}(y)+y_CS_{DAB}-y_DS_{CAB}(y),\nn\\
\e{6381}
where the arbitrary unphysical  functions $S_{ABC}$ satisfy 
\be
&&S_{ABC}(y)=-S_{ACB}(y),\quad S_{ABC}(y)+S_{BCA}(y)+S_{CAB}(y)=0,\nn\\
&&(y\cdot\dif+6)S_{ABC}(y)=0.
\e{6382}
Finally, we find from $\Box(y^A R_{ABCD})=0$  assuming the last relation in \rl{633} to be valid even off the hypercone
\be
&&\dif^AR_{ABCD}(y)=y_BS_{CD}(y)-y_C\tilde{S}_{DB}(y)+y_D\tilde{S}_{CB}(y),\nn\\&&S_{AB}(y)=\half y^CS_{CAB}(y),\quad \tilde{S}_{AB}(y)=\half y^CS_{ACB}(y),\nn\\&&\Rightarrow\;\;S_{AB}=\tilde{S}_{AB}-\tilde{S}_{BA}(y).
\e{639}

There are several problems with this $s=2$ theory. However, the main problem is the homogeneity condition \rl{638}, which in \rl{6371} requires
\be
&&(y\cdot\dif+1)H_{AB}(y)=0,
\e{6383}
in disagreement with the external field result in \rl{315}.
The complete Riemann tensor in terms of the symmetric metric tensor $G_{AB}(y)$ may be written as
\be
&&R_{ABCD}(y)=\half\biggl(\dif_A\dif_C G_{BD}(y)+\dif_B\dif_DG_{AC}(y)-\dif_B\dif_CG_{AD}(y)-\nn\\&&-\dif_A\dif_DG_{BC}(y)\biggr)+  G^{EF}(y)\biggl(\Gamma_{EAC}(y)\Gamma_{FBD}(y)-\Gamma_{EDA}(y)\Gamma_{FBC}(y)\biggr),\nn\\
\e{640}
where
\be
&&\Gamma_{ABC}(y)\equiv \half\biggl(\dif_BG_{AC}(y)+\dif_CG_{AB}(y)-\dif_AG_{BC}(y)\biggr).
\e{641}
The homogeneity condition in \rl{638} requires here
\be
&&(y\cdot\dif+1)G_{AB}(y)=0 \quad \Rightarrow\quad (y\cdot\dif-1)G^{AB}(y)=0
\e{643}
 in disagreement with the external field result in \rl{314}. Since the linearized Riemann tensor \rl{6371} follows from \rl{640} with $G_{AB}=\eta_{AB}+H_{AB}$ the results \rl{6383} and \rl{643} agree. The properties  in
\rl{643} are  bad  since they do not allow a constant flat metric in empty space ($G_{AB}=\eta_{AB}$). 

It is remarkable that the condition
\be
&&y^AR_{ABCD}(y)=0
\e{6431}
requires
\be
&&y\cdot\dif G_{AB}(y)=0,\quad y^AG_{AB}(y)=y_B
\e{642}
in agreement with the external field result in \rl{314}. This equation requires then
\be
&&y\cdot\dif H_{AB}(y)=0,\quad y^AH_{AB}(y)=0.
\e{6421}
in agreement with \rl{315}.

We are therefore convinced that the external field result is the correct result. In the next section we indicate what has to be changed in the considered particle models for $s\geq 2$.

 \setcounter{equation}{0}
\section{Homogeneity and the order of the field equations}
The appropriate homogeneity of the fields depends on the order of the field equations.
Consider the equation
\be
&&{\Box}^k\cF_{A_1\cdots A_{n};B_1\cdots B_n;C_1\cdots C_n;\cdots}(y)=0
\e{701}
for any positive integer $k$.
In abstract operator language this is given by the condition
\be
&&(p^2)^k|\cF\hb=0
\e{702} 
in terms of the general state vector \rl{606}. As before we also impose the hypercone condition
\be
&&y^2|\cF\hb=0.
\e{703}
Consistency between \rl{702} and \rl{703} requires
\be
&&[(p^2)^k, y^2]|\cF\hb=0,
\e{704}
which simplifies to
\be
&&(D-2k+2)|\cF\hb=0,
\e{705}
where $D$ is given by \rl{509}. Thus, instead of \rl{611} we have now the conditions
\be
&& y^2\cF_{\cdots }(y)=0,\nn\\
&&{\Box}^k \cF_{\cdots}(y)=0,\nn\\
&& (y\cdot\dif+d/2-k+2)\cF_{\cdots}(y)=0,
\e{706}
which reduce to
\be
&&\cF_{\cdots}(y)=\del(y^2)F_{\cdots}(y),
\e{707}
\be
&&{\Box}^kF_{\cdots}(y)=0,
\e{708}
\be
&&(y\cdot\dif+d/2-k)F_{\cdots}(y)=0.
\e{709}

\subsection{Implications for $s\geq2$}
In order to have the homogeneity
\be
&&y\cdot\dif G_{AB}(y)=0
\e{710}
as demanded by the external field result \rl{314}    and the condition that $G_{AB}(y)$ should contain the constant flat metric $\eta_{AB}$,  the Riemann tensor \rl{640} must satisfy   the homogeneity
\be
&&(y\cdot\dif+2)R_{ABCD}(y)=0
\e{711}
instead of \rl{638}.
For the original tensor field \rl{630} in $d=4$  this property of $R_{ABCD}$  requires 
\be
&&(y\cdot\dif+2)\cR_{ABCDEF}(y)=0
\e{712}
from \rl{632} and \rl{707}.
 Comparison beween \rl{712} and \rl{706} requires  then an equation of the type
\be
&&{\Box}^2\cR_{ABCDEF}(y)=0,
\e{713}
instead of \rl{631}. Notice then that \rl{614a} may not be imposed since \rl{614a} and \rl{614b} together imply the second order field equation. 

Now we do not have a complete particle theory for particles yielding tensor fields satisfying higher order equations. The above results are therefore heuristic. We may equally well imagine that we have complete theories for symmetric tensor fields like the external fields treated in section 4. The homogeneity \rl{710} implies then directly from \rl{708} and \rl{709} equations like 
\be
&&{\Box}^2G_{AB}(y)=0
\e{714}
in arbitrary dimensions. This indicates that  conformal gravity will enter here. This will be  confirmed  later.

Similarly from \rl{708} and \rl{709} the external field result \rl{322} implies equations of the order 
\be
&&{\Box}^sH_{A_1\cdots A_s}(y)=0,\quad s\geq 1,
\e{715}
where $s$ is the rank of the symmetric external tensor field. 

 \setcounter{equation}{0}
\section{Field theories in the conformal space}
So far fields have only entered as wave functions in the quantization of conformal particle models or as external fields in these models. Now we want to construct pure field theories formulated in the $d+2$ dimensional conformal space. The wave equations and the properties of the external fields in the previous sections will then serve as a guiding principle.

All treated manifestly conformally covariant particle models lead to a set of wave equations  where one equation is the restriction to the hypercone $y^2=0$. This equation has the form
\be
&&y^2\cF_{\cdots}(y)=0
\e{0801}
with the solution
\be
&&\cF_{\cdots}(y)=\del(y^2)F_{\cdots}(y).
\e{0802}
In the previous treatments we found that the external fields were more like the $F_{\cdots}$-fields. The peculiar delta function on the hypercone certainly makes the $F_{\cdots}$-fields  look more similar to the fields we usually handle as compared to $\cF_{\cdots}$. When we now develop a field theory we find it therefore natural to start from the $F_{\cdots}$-fields and   their properties. Now, $F_{\cdots}$ in \rl{0802} lives on the $d+1$ dimensional hypercone $y^2=0$. This means that the equation \rl{0801} reduces the dimension by one. One problem with this restriction is that  when dealing with the $F_{\cdots}$-fields one very easily may leave the hypercone by various manipulations. Following Dirac \cite{Dirac:1936} it is therefore better to let $F_{\cdots}(y)$
be defined in the $d+2$ dimensional conformal space and instead require  invariance under the gauge transformation
\be
&&F_{\cdots}(y)\;\;\ra\;\;F_{\cdots}(y)+y^2\tilde{F}_{\cdots}(y),
\e{0803}
where $\tilde{F}_{\cdots}(y)$ is an arbitrary function. The only restriction being that ${F}_{\cdots}(y)$ and $\tilde{F}_{\cdots}(y)$ must be smooth on the hypercone, \ie they must be  possible to  Taylor expand in $y^2$. The class of functions defined by the transformations \rl{0803} represents then one unique function on the hypercone $y^2=0$ but otherwise quite arbitrary. We shall refer to \rl{0803} as the special gauge transformation off the hypercone.

By means of \rl{0803} we may now develop a more geometrical field theory in $d+2$ dimensions. In the particle models we always have a condition of homogeneity. 
$\cF$ is \eg always required to have a definite degree of homogeneity, however, this implies that $F_{\cdots}$ only has a definite homogeneity up to $y^2$ terms. This is consistent with the gauge invariance under \rl{0803}. Now we believe (like Dirac) that we  always may restrict this arbitrariness by requiring a definite homogeneity in the whole conformal space (called strong homogeneity in the sequel). In the following treatments we shall therefore  always require  ${F}_{\cdots}(y)$ to have a  homogeneity of degree $n$ in the entire $d+2$ dimensional conformal space  (of course, chosen in accordance with previous results):
\be
&&(y\cdot\dif-n){F}_{\cdots}(y)=0.
\e{0804}
This implies that we partially fix the arbitrary functions $\tilde{F}_{\cdots}$ in \rl{0803} by the homogeneity condition
\be
&&(y\cdot\dif-n+2)\tilde{F}_{\cdots}(y)=0.
\e{0805}

The basic idea for the field theories in the $d+2$ dimensional conformal space to be defined below is that they will always be required to be invariant under the special gauge transformation \rl{0803}, which means that they are effectively defined on the $d+1$ dimensional hypercone. One consequence is then that one may  always fix the gauge by requiring
\be
&&{\dif\over\dif y^2}({F'}_{\cdots}(y))=0,\quad {F'}_{\cdots}(y)={F}_{\cdots}(y)+y^2\tilde{F}_{\cdots}(y),
\e{0806}
which means that $F'_{\cdots}$ is independent of $y^2$ and therefore is effectively defined in terms of $d+1$ coordinates. The homogeneity condition \rl{0804} may then be used to further reduce the field ${F'}_{\cdots}$ to $d$ dimensions as is described in appendix A. (Unfortunately, in the case of tensor fields to be described below we will be unable to choose the gauge \rl{0806} in this simple direct way  since we then have to further restrict $\tilde{F}_{\cdots}(y)$.)

\subsubsection{Scalar fields}
The Dirac quantization of the spinless conformal particle led to the equations \rl{203}. The condition \rl{0801} was solved by \rl{204}, and $\phi(y)$, which is the $F_{\cdots}$-field here, was required to satisfy equations \rl{205} on the hypercone $y^2=0$.  Imposing the corresponding strong homogeneity on $\phi(y)$ one may easily check that \rl{205} is invariant under the transformation
\be
&&\phi(y)\;\;\ra\;\;\phi(y)+y^2\tilde{\phi}(y)
\e{0807}
for arbitrary fields $\tilde{\phi}(y)$ only restricted by the homogeneity condition
\be
&&(y\cdot\dif+d/2-3)\tilde{\phi}(y)=0.
\e{08071}
The external scalar field in subsection 4.1 is also a field of this type and the formulation \rl{301} seems to be invariant under \rl{0807}.

\subsubsection{Spinor fields}
The $s=1/2$ model \rl{206} led to the field equations \rl{209} and \rl{211}. Two of the conditions were solved in \rl{212}, \ie
\be
&&\Psi(y)=\del(y^2)y\!\!\!\slash\psi(y),
\e{0808}
where then $\psi(y)$ has to satisfy \rl{213} on the hypercone $y^2=0$. Here we impose the strong homogeneity condition \rl{214}. One may easily check that \rl{0808} as well as \rl{213} and \rl{214} are invariant under the gauge transformations
\be
&&\psi(y)\;\;\ra\;\;\psi(y)+y\!\!\!\slash\chi(y),\quad (y\cdot\dif+d/2+1)\chi(y)=0.
\e{0809}
(This gauge transformation was not clearly stated by Dirac. However, it may be found in \eg \cite{Castell:1977}.)

\subsection{Tensor fields}
As soon as tensor fields are involved the situation becomes more complex. All  tensor fields satisfy transversality conditions like \rl{614c}.  For the $F_{\cdots}$-fields we have then transversality on the hypercone (called weak transversality in the sequel). Since we want geometrical field theories in the conformal space it is natural to impose strong transversality valid in the entire conformal space. This was \eg imposed for the external tensor fields in section 4. However, such a restriction causes complications what concerns the reduction to $d$ dimensions.  Below we explain these complications in the cases of $s=1$, and $s=2$ in $d=4$.

\subsubsection{$s=1$ fields}
The $s=1$ theory in section 6 for arbitrary dimensions $d$ yields the totally antisymmetric field, $F_{ABC\cdots}$, with $d/2+1$ indices from the basic condition \rl{0801}. Condition \rl{614d} is then solved by the expression
\be
&&F_{ABC\cdots}(y)=\sum_{antisym(ABC\cdots)}y_AF_{BC\cdots}(y),\nn\\
&&(y\cdot\dif+d/2-1)F_{ABC\cdots}(y)=0,\quad (y\cdot\dif+d/2)F_{BC\cdots}(y)=0,
\e{0810}
where $F_{BC\cdots}$ on the right-hand side is totally antisymmetric with $d/2$ indices. This latter field, which we also will call the intermediate $F$-field, may be transformed as follows
\be
&&F_{BC\cdots}(y)\quad\ra\quad F_{BC\cdots}(y)+\sum_{antisym(BC\cdots)}y_BU_{CD\cdots}(y),\nn\\
&&(y\cdot\dif+d/2+1)U_{CD\cdots}(y)=0,
\e{0811}
without affecting the original field in \rl{0810}. $U_{CD\cdots}(y)$ are arbitrary antisymmetric functions with $d/2-1$ indices. The transversality condition \rl{614c} requires
\be
&&y^AF_{ABC\cdots}(y)=y^2F'_{BC\cdots}(y),
\e{0812}
where consistency requires
\be
&&y^BF'_{BC\cdots}(y)=0.
\e{0813}
Equations \rl{0810}, \rl{0811} and \rl{0813} implies
\be
&&F_{BC\cdots}(y)=F'_{BC\cdots}(y)+\sum_{antisym(BC\cdots)}y_BU_{CD\cdots}(y).
\e{0814}
Condition \rl{614b} may be solved by the expression
\be
&&F_{BCD\cdots}(y)=\sum_{antisym(BCD\cdots)}(\dif_BA_{CD\cdots}(y)),\nn\\
&&(y\cdot\dif+d/2-1)A_{CD\cdots}(y)=0,
\e{0815}
which implies 
\be
&&y^BF_{BCD\cdots}(y)=-\sum_{antisym(BCD\cdots)}\dif_C(y^BA_{BD\cdots}(y)).
\e{0816}
In order for $F_{BCD\cdots}$ in \rl{0815} to have the general form \rl{0814} we have to assume that $A_{BD\cdots}$ satisfies weak transversality
\be
&&y^BA_{BC\cdots}(y)=y^2B'_{CD\cdots}(y)\quad\Rightarrow\nn\\
&&(y\cdot\dif+d/2)B'_{CD\cdots}(y)=0,\quad y^CB'_{CD\cdots}(y)=0,
\e{0817}
where $B'_{CD\cdots}$ is totally antisymmetric with $d/2-2$ indices. 
Combining \rl{0814}, \rl{0815}, \rl{0816} and \rl{0817} we find that \rl{0815} requires $U_{CD\cdots}$ in \rl{0814} to be of the form
\be
&&U_{CD\cdots}(y)=\sum_{antisym(CD\cdots)}(\dif_CB'_{D\cdots}(y)),
\e{08171}
 and that $F'_{BCD\cdots}$ in  \rl{0814}  is of the form \rl{0815} with $A_{BC\cdots}$ replaced by $A'_{BC\cdots}$ satisfying strong transversality
\be
&&y^BA'_{BC\cdots}(y)=0,
\e{0818}
This allows us to set
\be
&&A_{BC\cdots}(y)=A'_{BC\cdots}(y)+\sum_{antisym(BCD\cdots)}(y_BB'_{CD\cdots}(y))
\e{0819}
Notice that we may replace $B'_{CD\cdots}$ by $B_{CD\cdots}$ without affecting $A_{BC\cdots}$ by the expression
\be
&&B_{BC\cdots}(y)=B'_{BC\cdots}(y)+\sum_{antisym(BCD\cdots)}(y_B\phi'_{CD\cdots}(y)),\nn\\
&&(y\cdot\dif+d/2+1)\phi'_{CD\cdots}(y)=0,\quad y^C\phi'_{CD\cdots}(y)=0,
\e{08191}
where $\phi'_{CD\cdots}$ is totally antisymmetric with $d/2-3$ indices. $B_{BC\cdots}$ satisfies then weak transversality
\be
&&y^BB_{BCD\cdots}(y)=y^2\phi'_{CD\cdots}(y).
\e{08192}
One may then proceed and replace $\phi'_{CD\cdots}$ in \rl{08191} by $\phi_{CD\cdots}$ satisfying weak transversality by a similar expression etc. Anyway transformations of the type
\be
&&A_{BC\cdots}(y)\quad\ra\quad A_{BC\cdots}(y)+\sum_{antisym({BC\cdots})}(y_BK_{C\cdots}(y)),
\e{08201}
 are obviously gauge transformations in the theory if  $K_{C\cdots}$ is totally antisymmetric satisfying at least weak transversality. In addition we have of course the standard gauge invariance of the intermediate $F$-field \rl{0815} under the transformations
 \be
 &&A_{CD\cdots}(y)\quad\ra\quad A_{CD\cdots}+\sum_{antisym(CD\cdots)}\dif_C\Lambda_{D\cdots}(y),\nn\\
 &&(y\cdot\dif+d/2-2)\Lambda_{D\cdots}(y)=0
 \e{082011}
 for arbitrary antisymmetric functions $\Lambda_{D\cdots}$

Consider now the special gauge transformations off the hypercone. For the basic elementary fields $A_{BC\cdots}$ with $d/2-1$ indices this transformation is of the form
\be
&&A_{BC\cdots}(y)\quad\ra\quad A_{BC\cdots}(y)+y^2\tilde{A}_{BC\cdots}(y),\nn\\
&&(y\cdot\dif+d/2+1)\tilde{A}_{BC\cdots}(y)=0.
\e{08202}
This implies for the intermediate $F$-field with $d/2$ indices
\be
&&F_{BCD\cdots}(y)\quad\ra\quad F_{BCD\cdots}(y)+y^2\tilde{F}_{BCD\cdots}(y)+\nn\\
&&+\sum_{antisym(BCD\cdots)}2(y_B\tilde{A}_{CD\cdots}(y))
\e{08203}
from \rl{0815}. The last terms are of the form of the last terms in \rl{0811} ($U_{CD\cdots}=2\tilde{A}_{CD\cdots}$). The original $F_{ABC\cdots}$-fields with $d/2+1$ indices transforms therefore exactly like the ${A}_{BC\cdots}$-fields:
\be
&&F_{ABC\cdots}(y)\quad\ra\quad F_{ABC\cdots}(y)+y^2\tilde{F}_{ABC\cdots},
\e{08204}
where $\tilde{F}_{ABC\cdots}$ is exactly the same expression as $F_{ABC\cdots}$ but with $A_{BC\cdots}$ replaced by $\tilde{A}_{BC\cdots}$.

Due to their large invariances it looks like the original field in \rl{0810} should be particularly useful. However, this is not the case. From \rl{0812} it follows that this field always satisfy weak transversality even when the intermediate $F$-field and the elementary $A$-field satisfy strong transversality. Another peculiar property is that the divergence of these fields are expressed in terms of the $F'$-fields (see section 6 and eq.\rl{082092}  below). Furthermore, we have not found any natural way that they may enter the Lagrangian theory to be treated in the next section.

 In order to define a maximally geometric field theory  it is natural to impose strong transversality. We have seen that the gauge invariances of the theory always allow us to choose the $A$-field to be strongly transverse,
 \be
 &&y^BA_{BC\cdots}(y)=0,
 \e{08205}
 which implies
  \be
 &&y^BF_{BC\cdots}(y)=0,
 \e{08206}
 for the intermediate $F$-field with $d/2$ indices. With these conditions the field theory becomes more geometric. The price we pay for these conditions is that the gauge degrees of freedom of the intermediate $F$-field now is fixed. The condition \rl{08205} does not allow for gauge transformations of the form \rl{08201}. What is more serious is that the special gauge transformations off the hypercone \rl{08202} also have to be restricted. In fact, the condition \rl{08205} forces us to impose
 \be
 &&y^B\tilde{A}_{BC\cdots}(y)=0
 \e{08207}
 in \rl{08202}.  Unfortunately, this restriction is then  too strong to allow for the gauge \rl{0806}. In fact, this is obvious since \rl{08205} does not allow for a solution of the form \rl{0806}. What we have to do then is to remember that the solution is of the form 
\be
&&A_{BC\cdots}(y)=V_{BC\cdots}(y)-\sum_{antisym(BC\cdots)}(y_B\phi_{CD\cdots}(y),
\e{08208}
from \rl{0819} where both $V_{BC\cdots}$ and $\phi_{CD\cdots}$ satisfy weak transversality,
\be
&&y^BV_{BC\cdots}(y)=y^2\phi_{CD\cdots}(y),\quad    y^C\phi_{CD\cdots}=y^2\Lambda_{D\cdots}(y).
\e{08209}
The fields $V_{BC\cdots}$, $\phi_{CD\cdots}$ and $\Lambda_{D\cdots}$ are totally antisymmetric with $d/2-1$, $d/2-2$ and $d/2-3$ indices, with the homogeneities
\be
&&(y\cdot\dif+d/2-1)V_{BC\cdots}(y)=0,\quad      (y\cdot\dif+d/2)   \phi_{CD\cdots}(y)  =0,\nn\\
&&(y\cdot\dif+d/2+1)\Lambda_{D\cdots}(y)=0.
\e{082091}
For the fields $V_{BC\cdots}$ and  $\phi_{CD\cdots}$ in \rl{08208} we may apply unrestricted gauge transformations off the hypercone which means that for these fields we may choose the gauge \rl{0806}. The $y^2$-dependence in $A_{BC\cdots}$ is then isolated in the explicit $y$'s in \rl{08208}. (See appendix B where the reduction for $d=4$ is explicitly performed.)

Consider now the equations of motion. For the original field in \rl{0810} we find from \rl{614a} and \rl{6131} on the hypercone
\be
&&\dif^AF_{ABC\cdots}(y)=2F_{BC\cdots}(y),\nn\\
&&\Box F_{ABC\cdots}(y)=0.
\e{082092}
From the expression in \rl{0810} we find for the intermediate $F$-field 
the equations \rl{6291} on the hypercone which are given by
\be
&&\Box F_{ABC\cdots}(y)=\sum_{antisym(ABC\cdots)}y_AS_{BC\cdots}(y), \nn\\&&\dif^AF_{ABC\cdots}(y)=\sum_{antisym(BC\cdots)}y_BS_{CD\cdots}(y),\nn\\
&& (y\cdot\dif+d/2+3)S_{BC\cdots}(y)=0,\nn\\
&&(y\cdot\dif+d/2+2)S_{CD\cdots}(y)=0,\nn\\
&&S_{CD\cdots}(y)=\half y^BS_{BCD\cdots}(y),
\e{082093}
where $S_{ABC\cdots}$ are arbitrary  totally antisymmetric functions. These equations are consistent with the strong transversality \rl{08206}, and the relations between the equations follow from the consistency condition $\Box (y^AF_{ABC\cdots})=0$. Notice also that the second equation satisfies the consistency condition $\dif^A(y^BF_{ABC\cdots})=0$. The $S$-functions do not affect the original fields and are therefore unphysical. Only in this form are the equations invariant under the special gauge transformations off the hypercone. The transformations \rl{08202} only affect the $S$-functions. We find
\be
&&S_{BC\cdots}(y)\quad\ra\quad S'_{BC\cdots}(y)=S_{BC\cdots}(y)+2\Box \tilde{A}_{BC\cdots}(y),\nn\\
&&S_{CD\cdots}(y)\quad\ra\quad S'_{CD\cdots}(y)=S_{CD\cdots}(y)-2\dif^B\tilde{A}_{BC\cdots}(y),\nn\\
&&S'_{CD\cdots}(y)=\half y^BS'_{BCD\cdots}(y).
\e{082094}
The $S$-functions reduce the number of equations for the intermediate field $F_{ABC\cdots}$ and their values should be ignored at the end since they are unphysical.

\subsection{$s=2$ fields in $d=4$}
In section 6.2 and in section 7 we saw that the $s=2$ quantum theory in $d=4$ of the conformal particle was not quite right. The problems were mainly due to the homogeneity and the order of the equations as explained in section 7. Still we found that the Riemann tensor seems to be very natural and should be involved in the correct theory. To start with we assume therefore that the linearized Riemann tensor \rl{6371} is relevant and important. It is
\be
 &&R_{ABCD}(y)=\half\biggl(\dif_A\dif_C H_{BD}(y)+\nn\\&&+\dif_B\dif_DH_{AC}(y)-\dif_B\dif_CH_{AD}(y)-\dif_A\dif_DH_{BC}(y)\biggr),
 \e{0821}
 where $H_{AB}$ is the elementary field. It is symmetric and satisfies the strong homogeneity condition
 \be
 &&y\cdot\dif H_{AB}(y)=0\quad \Rightarrow\quad (y\cdot\dif+2)R_{ABCD}(y)=0.
 \e{0822}
 The special gauge transformation off the hypercone
 \be
 &&H_{AB}(y)\;\;\ra\;\;H_{AB}(y)+y^2\tilde{H}_{AB}(y),\nn\\
 &&(y\cdot\dif+2)\tilde{H}_{AB}(y)=0,
 \e{0823}
 implies then for \rl{0821}
 \be
 &&R_{ABCD}(y)\;\;\ra\;\;R_{ABCD}(y)+y^2\tilde{R}_{ABCD}(y)+\nn\\
 &&+\eta_{AC}\tilde{H}_{BD}(y)+\eta_{BD}\tilde{H}_{AC}(y)-\eta_{BC}\tilde{H}_{AD}(y)-\eta_{AD}\tilde{H}_{BC}(y)+\nn\\
 &&+y_A(\dif_C\tilde{H}_{BD}(y)-\dif_D\tilde{H}_{BC}(y))+y_B(\dif_D\tilde{H}_{AC}(y)-\dif_C\tilde{H}_{AD}(y))+\nn\\
 &&+y_C(\dif_A\tilde{H}_{BD}(y)-\dif_B\tilde{H}_{AD}(y))+y_D(\dif_B\tilde{H}_{AC}(y)-\dif_A\tilde{H}_{BC}(y)).
 \e{0824}
 Also here there are problems with the theory in section 6.2 since
these transformation properties do {\em not} imply that $R_{ABCDEF}$ in section 6.2 transforms like
 \be
 &&R_{ABCDEF}(y)\;\;\ra\;\;R_{ABCDEF}(y)+y^2\tilde{R}_{ABCDEF}(y),
 \e{0825}
 where $\tilde{R}_{ABCDEF}(y)$ is $R_{ABCDEF}(y)$ with $H_{AB}$ replaced by $\tilde{H}_{AB}$. For a correct theory this is what we should expect. Although the $y$-terms are of the form \rl{6321} with
 \be
 &&U_{BCD}(y)=\dif_C\tilde{H}_{BD}(y)-\dif_D\tilde{H}_{BC}(y),
 \e{08251}
 which means that $R_{ABCDEF}$ is invariant under this part of \rl{0824}, 
the $\eta$-terms in \rl{0824} yield additional terms on the right hand side of \rl{0825}. These terms are due to the fact that \rl{0821} are of second order in the derivatives. Although we have not found the correct $s=2$ theory and the correct underlying conformal particle model, we proceed now with the assumption that the linearized Riemann tensor \rl{0821} is a relevant intermediate field and that \rl{0824} is an invariance transformation on the hypercone. In the next section we shall show that this theory exists and also determine its form within the action formalism.
 
One may notice that  the  tensor
 \be
 &&W_{ABCD}(y)=\al R_{ABCD}(y)+\nn\\&&+\beta (\eta_{AC}R_{BD}(y)+\eta_{BD}R_{AC}(y)-\eta_{BC}R_{AD}(y)-\eta_{AD}R_{BC}(y))+\nn\\&&+\ga(\eta_{AC}\eta_{BD}-\eta_{BC}\eta_{AD})R(y),
 \e{0826}
 where we have introduced the linearized Ricci tensor and curvature scalar given by
 \be
 &&R_{BD}=\eta^{AC}R_{ABCD}(y),\quad R=\eta^{AB}R_{AB}(y),
 \e{08261}
 is insensitive to the $\eta$-terms in \rl{0824} for
 \be
 &&\al+4\beta=0,\quad \beta+5\ga=0.
 \e{0827}
 In fact, for $\al=1$ $W_{ABCD}$ is then the Weyl tensor in six dimensions. This is an indication that we are here driving towards conformal gravity. Notice, however, that $W_{ABCD}$ is not the invariant tensor we are looking for since it is not invariant under the $y$-terms in \rl{0824}.

 Assume now that weak transversality is allowed. For $H_{AB}$ this requires
 \be
 &&y^AH_{AB}(y)=y^2V_B(y),\quad (y\cdot\dif+1)V_A(y)=0.
 \e{08271}
We also assume weak transversality for $V_A$,
  \be
 &&y^AV_A(y)=y^2\phi(y)\quad (y\cdot\dif+2)\phi(y)=0.
 \e{082711}
We may then  replace \rl{08271} by the strong relation
 \be
 &&y^AH'_{AB}(y)=0,
 \e{08272}
 where
 \be
 &&H'_{AB}(y)=H_{AB}(y)-y_AV_B(y)-y_BV_A(y)+y_Ay_B\phi(y),
 \e{08273}
 which also may be written as
  \be
 &&H'_{AB}(y)=H_{AB}(y)-y_AV'_B(y)-y_BV'_A(y)-y_Ay_B\phi(y),
 \e{08274}
 where
 \be
 &&V'_A(y)=V_A(y)-y_A\phi(y),\quad y^AV'_A(y)=0.
 \e{08275}
 Performing the transformation $H_{AB}\;\ra\;H'_{AB}$ in the linearized Riemann tensor \rl{0821} we find
  \be
 &&R_{ABCD}(y)\;\;\ra\;\;R'_{ABCD}(y)=R_{ABCD}(y)+\nn\\
 &&+\half\biggl(\eta_{AC}M_{BD}(y)+\eta_{BD}M_{AC}(y)-\eta_{BC}M_{AD}(y)-\eta_{AD}M_{BC}(y)+\nn\\
 &&+y_AL_{BCD}(y)-y_BL_{ACD}(y)+ y_CL_{DAB}(y)-y_DL_{CAB}(y)\biggr),\nn\\
 &&M_{AB}(y)\equiv \dif_AV_B(y)+\dif_BV_A(y)-(y_A\dif_B+y_B\dif_A)\phi(y)-\eta_{AB}\phi(y),\nn\\
 &&L_{ABC}(y)\equiv \dif_AF_{BC}+\half L_{BC}(y)\dif_A\phi(y),\nn\\
 && F_{AB}(y)=\dif_AV_B(y)-\dif_BV_A(y).
 \e{08276}
 This is  very similar in structure to \rl{0824}. The $y$-terms are of the form \rl{6321} and the $\eta$-terms of the form of the $\eta$-terms in \rl{0824}.   We assume therefore that this is an invariance transformation in the theory.   This gauge invariance we may fix by imposing the strong transversality
 \be
 &&y^AH_{AB}(y)=0\quad\Rightarrow\quad y^AR_{ABCD}(y)=0.
 \e{0832}
 In fact, \rl{0822} and \rl{0832} must be strongly valid by  geometrical reasons from \rl{642}  and the external field results \rl{315}.
However, since the theory we are looking for no longer is   invariant under transformations of the type $H_{AB}\;\ra\;H'_{AB}$ there is a technical difficulty to go down to $d$ dimensions.

  The basic special gauge invariance under \rl{0823} is restricted by the conditions \rl{0832}. We have
  now 
  invariance under \rl{0823}, where $\tilde{H}_{AB}$ satisfies
 \be
 &&(y\cdot\dif+2)\tilde{H}_{AB}(y)=0,\quad y^A\tilde{H}_{AB}(y)=0,
 \e{0834}
which no longer allow for the gauge choice \rl{0806}.
 The strong condition \rl{0832} makes it also impossible to choose $H_{AB}$ to be independent of $y^2$. Only the form $H'_{AB}$ in \rl{08273} is such that $H$, $V$ and $\phi$ may be chosen to be independent of $y^2$. All $y^2$ dependence is then isolated in the explicit $y$'s  (see appendix C).

Since the $s=2$ theory in section 6 is not quite the theory we are looking for, the equations \rl{6372}-\rl{639} are irrelevant here. In the next section we derive the correct equations of motion.
 Finally it should be mentioned that the linearized Riemann tensor \rl{0821} is invariant under the ordinary  gauge transformation
 \be
 &&H_{AB}(y)\;\;\ra\;\;H_{AB}(y)+\dif_AA_B(y)+\dif_BA_A(y),\nn\\&&y^AA_A(y)=0,\quad (y\cdot\dif-1)A_A(y)=0,
 \e{0828}
which is consistent with the strong transversality \rl{0832}.

 \setcounter{equation}{0}
\section{Lagrangians and the action principle for manifestly conformal fields}
In this section we propose a powerful consistent action principle for manifestly conformal fields. Actions for manifestly conformal field theories were first seriously treated in \cite{Budini:1979}. However, in these actions the measure contained a non-covariant delta function. No manifestly conformally covariant equations were therefore possible to derive. In \cite{Marnelius:1980} one of us proposed the use of actions with an invariant measure on the hypercone. (Two examples  in $d=4$ were given:  free spin one-half and free spin one.) This formulation is  here generalized to arbitrary fields in arbitrary dimensions and in addition we  give a precise prescription how they are  defined and how they are to be treated.  We expect this principle to be a powerful means to deal with  conformal field theories. Some examples are given below.

The main obstacle to set up an action for the manifestly conformal fields is that they satisfy more than one equation. Of course we could try to introduce extra fields,  Lagrange multipliers,  for some of the equations. However, here we shall avoid any introduction of extra fields and simply demand that  the given fields satisfy some of the required conditions leaving only one field equation to be derived from the action. 

The actions to be considered here will all have the form
\be
&&A=\int dy\del(y^2)\cL(y),
\e{801}
where $dy$ is the natural flat measure on the conformal space involving  the $d+2$ coordinates $y^A$. The Lagrangian $\cL(y)$ is local in the fields and is required to be a scalar under $SO(d,2)$-transformations. The delta function $\del(y^2)$ is inserted in order for the  action to be defined on the $d+1$-dimensional hypercone  $y^2=0$. In fact, we  require the action to produce equations valid only on the hypercone in consistency with previous equations. 
We also require that the action \rl{801} does not depend on the length scale of the coordinates $y^A$. In order to secure this we require that the Lagrangian density $\cL(y)$ is homogeneous with the degree of homogeneity $-d$, \ie we require the strong relation
\be
&&(y\cdot\dif+d)\cL(y)=0.
\e{802}
Now, we demand that each manifestly conformal field in the action satisfies a definite strong homogeneity condition in $d+2$ dimensions. Condition \rl{802} determines then the possible forms of the Lagrangians $\cL(y)$. (This corresponds to dimensional counting in $d$ dimensions.)  In addition,   the manifestly conformal fields in the action must satisfy a strong transversality condition. Thus, any field $a_{ABC\cdots}(y)$ in \rl{801} satisfies typically the conditions
\be
&&(y\cdot\dif-n)a_{ABC\cdots}(y) =0,\quad y^Aa_{ABC\cdots}(y)=0,\nn\\&& y^Ba_{ABC\cdots}(y)=0,\quad y^Ca_{ABC\cdots}(y)=0, \quad {\rm etc.},
\e{8021}
where the constant $n$ is the degree of homogeneity of the field $a_{ABC\cdots}(y)$. A further condition is that it must be allowed to view all fields in $\cL(y)$ as coming from  equations like $y^2\cF(y)=0$, \ie we must require that all fields in $\cL(y)$ are possible to restrict to the hypercone directly or indirectly as explained in section 8.  This condition is satisfied if we require the action to be invariant under the special gauge transformations given in the previous section for each involved field.  These gauge transformations are here of the form 
\be
&&a_{ABC\cdots}(y) \;\longrightarrow\;a'_{ABC\cdots}(y) = a_{ABC\cdots}(y) +y^2\tilde{a}_{ABC\cdots}(y),
\e{80211}
where $\tilde{a}_{ABC\cdots}(y)$ are arbitrary functions only restricted by the conditions 
\be
&&(y\cdot\dif-n+2)\tilde{a}_{ABC\cdots}(y) =0,\quad y^A\tilde{a}_{ABC\cdots}(y)=0,\nn\\
&& y^B\tilde{a}_{ABC\cdots}(y)=0,\quad y^C\tilde{a}_{ABC\cdots}(y)=0, \quad {\rm etc.},
\e{80212}
which makes $a'_{ABC\cdots}(y)$  satisfy \rl{8021}.  The transversality conditions on $\tilde{a}_{ABC\cdots}$ requires the indirect reduction of $a_{ABC\cdots}(y)$ to $d$ dimensions.

When deriving equations from the action \rl{801} we  require that the variations are consistent with the imposed conditions \rl{8021}, \ie we require also the variations $\del a$  to satisfy \rl{8021}. In the procedure to derive the equations we always drop all total divergences even though they  involve the delta function $\del(y^2)$.  Under all these prescriptions we find that all derived equations are of the form
\be
&&\del(y^2)(\cdots)=0,
\e{8022}
\ie the derived equations are always valid only on the hypercone $y^2=0$. Furthermore,  in all examples we consider, which are consistent with the above requirements,  the derived equations are consistent with the imposed conditions of the form \rl{8021} on the hypercone.  We illuminate now the procedure by means of  some explicit examples.

\subsection{Scalar field theories}
We quantized the spinless conformal particle in section 3 and found the three equations \rl{203}. By means of \rl{204} they were then reduced to the two equations \rl{205}, one of which is the homogeneity condition
\be
&&(y\cdot\dif+d/2-1)\phi(y)=0,
\e{803}
here viewed as a strong relation.
This suggests the following form of the Lagrangian density $\cL(y)$:
\be
&&\cL_0(y)=\half \phi(y)\Box\phi(y),
\e{804}
which satisfies the condition \rl{802} for any dimension $d$. Furthermore, it is invariant under  gauge transformations of the form \rl{80211}. A variation of the corresponding action \rl{801} for fields satisfying the homogeneity condition \rl{803} produces then the equation
\be
&&\Box\phi(y)=0
\e{805}
on the hypercone $y^2=0$. Also \rl{805} is invariant under the gauge transformations \rl{80211}. However,  notice that $\cL_0=1/2\dif_A\phi\dif^A\phi$ does not reproduce \rl{805} for $d>2$.   In fact, the corresponding action is not invariant under the gauge transformations \rl{80211} and does not yield an equation on the hypercone $y^2=0$ except for $d=2$.

We may also add interaction terms to the free Lagrangian \rl{804}, like polynomials in the scalar fields. However, this is possible only in $d\leq6$ due to the condition \rl{802}. We have arbitrary self-interactions in $d=2$ and self-interactions of the type
\be
&&\cL(y)=\cL_0(y)+\phi^4(y)\quad {\rm in}\quad d=4,\nn\\
&&\cL(y)=\cL_0(y)+\phi^3(y)\quad {\rm in}\quad d=6.
\e{806}
This is also in agreement with the external field result \rl{303}-\rl{304}. (Lagrangians for scalar fields were also discussed in \cite{Budini:1979}.)

\subsection{Spinor field theories}
The $s=1/2$ particle model produced the equations \rl{209}  and \rl{211}. They were partly solved by \rl{212} leaving \rl{213}. In particular we have the homogeneity condition \rl{214}
\be
&&(y\cdot\dif+d/2)\psi(y)=0.
\e{807}
Imposing this as a strong relation 
we are here led to the Lagrangian density \cite{Marnelius:1980}
\be
&&\cL_0(y)=\half \bar{\psi}(y)y\!\!\!\slash\dif\!\!\!\slash \psi(y),\quad  \bar{\psi}(y)\equiv\psi^{\dag}\Gamma^0\Gamma^{d+2},
\e{808}
which from \rl{807} satisfies the condition \rl{802}. The corresponding action produces  the equation
\be
&&y\!\!\!\slash\dif\!\!\!\slash\psi(y)=0
\e{809}
on the hypercone $y^2=0$ provided the field $\psi(y)$ in \rl{808} satisfies \rl{807}. This agrees with \rl{213}. There is no allowed self-interaction without derivatives. The action to \rl{808} and the equations \rl{809} are invariant under the gauge transformation \rl{0809}. The Lagrangian \rl{808} and its reduction to  $d=4$ was given in \cite{Marnelius:1980}. (The spinor Lagrangians in \cite{Budini:1979} are not consistent with our conditions.)

\subsection{$s=1$ field theories}
The $s=1$ models in section 6 led naturally to antisymmetric tensor fields with $d/2+1$ indices. However, we have found no good actions for these fields. It turned out, however, that these fields could always be expressed in terms of 
an antisymmetric tensor field, $F_{\cdots}(y)$, with $d/2$ indices satisfying the equations \rl{6291}/\rl{082093}. Condition \rl{614b} could then be solved by expressing $F_{\cdots}(y)$ in terms of a sum of antisymmetric fields $A_{\cdots}(y)$ with $d/2-1$ indices differentiated once. Taking the homogeneity condition in \rl{0810} in the strong sense, the following Lagrangian density which is consistent with the condition \rl{802} suggests itself (the normalization is just a choice):
\be
&&\cL_0(y)={1\over 4}F_{ABC\cdots}(y)F^{ABC\cdots}(y),
\e{810}
where $F_{\cdots}$ is given by the expression \rl{0815}, \ie 
\be
&&F_{ABC\cdots}(y)=\sum_{antisym (ABC\cdots)}\dif_AA_{BC\cdots}(y).
\e{811}
(For $d=2,4,6$ we have the  expressions \rl{6143}, \rl{617}, and \rl{627}.) The corresponding action to \rl{810} yields naively the equations 
\be
&&\dif^AF_{ABC\cdots}(y)=0
\e{812}
on the hypercone $y^2=0$ provided the $A$-fields in \rl{811} satisfy the strong  homogeneity condition in \rl{0815}, \ie
\be
&&(y\cdot\dif+d/2-1)A_{BC\cdots}(y)=0
\e{813}
and the strong transversality condition
\be
&&y^AF_{ABC\cdots}(y)=0,
\e{814}
which requires
\be
&&y^BA_{BC\cdots}(y)=0.
\e{815}
Condition \rl{814} makes the action to \rl{810} invariant under the special gauge transformations \rl{80211} for $A_{BC\cdots}$ given by
\be
&&A_{BC\cdots}(y)\quad\ra\quad A_{BC\cdots}(y)+y^2\tilde{A}_{BC\cdots}(y),\nn\\
&&(y\cdot\dif+d/2+1)\tilde{A}_{BC\cdots}(y)=0,\quad y^B\tilde{A}_{BC\cdots}(y)=0.
\e{81505}
However, the naive equations \rl{812} are not  invariant under these gauge transformations. 
We find 
\be
&&\dif^AF_{AB\cdots}(y)\quad\ra\quad\dif^AF_{AB\cdots}(y)+\sum_{antisym(BC\cdots)} 2y_B\dif^A\tilde{A}_{AC\cdots}(y)\nn\\
\e{8151}
on the hypercone $y^2=0$.
In fact, the equations \rl{812} are incorrect since
the conditions \rl{815} also restrict the variations of $A_{BC\cdots}$. The correct equations from the action are instead
\be
&&\dif^AF_{ABCD\cdots}(y)=\sum_{antisym(BC\cdots)}y_BS_{CD\cdots}(y)
\e{8152}
on the hypercone $y^2=0$,
where $S_{BC\cdots}$ are arbitrary unphysical  functions. It is remarkable that exactly the same equations appear here as in eq.\rl{6291} in section 6 (and in eq.\rl{082093} in section 8). From the self-consistency of \rl{8152} it follows that  $S_{BC\cdots}$    is totally antisymmetric and  satisfies the weak relations
\be
&&(y\cdot\dif+d/2+2)S_{ABC\cdots}(y)=y^2f_{ABC\cdots}(y),\nn\\&&y^AS_{ABC\cdots}(y)=y^2g_{ABC\cdots}(y),
\e{8153}
in agreement with \rl{6291}/\rl{082093}.
$S_{ABC\cdots}$ reduces the number of equations for $A_{BC\cdots}$ (which compensates for \rl{815}). 
The correct equations of motion must be gauge invariant since the actions are gauge invariant. However, due to the restricted variations the equations are not quite tensor equations. They are tensor equations modulus terms involving
 the arbitrary unphysical functions $S_{CD\cdots}$.  As we pointed out in the previous section the special gauge transformations \rl{81505} only affect   $S_{CD\cdots}$:
\be
&&S_{CD\cdots}(y)\quad\Rightarrow\quad S'_{CD\cdots}(y)=S_{CD\cdots}(y)-2\dif^A\tilde{A}_{ABC\cdots}(y),
\e{8154}
where both $S_{CD\cdots}$ and $S'_{CD\cdots}$ are equally arbitrary. Since the  equations are 
such that the values of  $S_{CD\cdots}$ may be ignored, they are  not affected by \rl{81505} and are therefore gauge invariant. See also appendix B for further clarifications.

In $d=4$ and $d=6$ we may add polynomial interaction terms of the following types (in $d=2$ we have a scalar theory)
\be
&&\cL(y)=\cL_0(y)+A^4(y)+A^2(y)\dif A(y)\quad {\rm in}\quad d=4,\nn\\
&&\cL(y)=\cL_0(y)+A^3(y)\quad {\rm in}\quad d=6.
\e{816}
In $d=4$ we have \eg manifestly conformal Yang-Mills theories (previously treated in the manifest language in \cite{Ichinose:1985xb} and recently in \cite{Bars:2006dy}). It is clear that one has to check that derivative terms are invariant under the gauge transformations \rl{80211}.  In fact, the manifestly conformal Yang-Mills action is invariant.
 We may also give interaction terms which  combine scalar, spinor,  vector and general $s=1$ fields. In this way we may  \eg obtain the massless standard model within this manifestly conformal language. Examples of possible interaction terms are
 \be
 &&\bar{\psi}(y)y\!\!\!\slash\Gamma^A\psi(y)A_A(y),\quad
 \bar{\psi}(y)y\!\!\!\slash\psi(y)\phi(y),
 \e{8161}
 where the first one is consistent for any $d$ while the second one is consistent only for $d=4$.
 
 \subsection{Linear manifestly conformal gravity in $d=4$}
 In order to have the correct homogeneity \rl{710} for the metric tensor the Riemann tensor must satisfy \rl{711}.
 A Lagrangian satisfying \rl{802} in $d=4$ must then have the form
 \be
 &&\cL_0(y)=\al R_{ABCD}(y)R^{ABCD}(y)+\beta R_{AB}(y)R^{AB}(y)+\ga R^2(y),
 \e{817}
 where $\al$, $\beta$, and $\ga$ are real constants. In order to obtain linear equations from \rl{817} we have to use the linearized Riemann tensor \rl{0821}, \ie
 \be
 &&R_{ABCD}(y)=\half\biggl(\dif_A\dif_C H_{BD}(y)+\nn\\&&+\dif_B\dif_DH_{AC}(y)-\dif_B\dif_CH_{AD}(y)-\dif_A\dif_DH_{BC}(y)\biggr),
 \e{819}
 from which we obtain 
 \be
 &&R_{AB}(y)=\eta^{CD}R_{ACBD}(y)=\half\biggl(\dif_A\dif_BH^C_{\;\;C}(y)+\dif_C\dif^CH_{AB}(y)-\biggr.\nn\\&&\biggl.-\dif_A\dif_C H^C_{\;\;B}(y)-\dif_B\dif_CH^C_{\;\;A}(y)\biggr),\nn\\
 &&R(y)=\eta^{AB}R_{AB}(y)=\dif_A\dif^AH^B_{\;\;B}(y)-\dif_A\dif_B H^{AB}(y).
 \e{820}
Consistency requires $H_{AB}(y)$  to satisfy
  \be
 &&y\cdot\dif H_{AB}(y)=0,\quad y^AH_{AB}(y)=0,
 \e{821}
 in the strong sense which also were argued for in sections 6 and  8.
 It remains to investigate whether or not the corresponding action to \rl{817} is invariant under the special gauge transformations \rl{80211} which here have the form
 \be
 &&H_{AB}(y)\;\;\ra\;\;H'_{AB}(y)=H_{AB}(y)+y^2\tilde{H}_{AB}(y),
 \e{8201}
 where $\tilde{H}_{AB}$ is a symmetric field satisfying
 \be
 &&(y\cdot\dif+2)\tilde{H}_{AB}(y)=0,\quad y^A\tilde{H}_{AB}(y)=0,
 \e{8202}
 which is required in order for $H'_{AB}$ to satisfy \rl{821}.

 We find now that the action to \rl{817} is invariant under \rl{8201} provided
  \be
 &&\al=3\ga,\quad \beta=-6\ga.
 \e{822}
 (The details of these calculations are similar to the corresponding calculations in section 12 (see \rl{1120}-\rl{1122}).)
 Independently, we find that \rl{817} yields equations on the hypercone $y^2=0$, \ie equations without  terms with derivatives of the delta function $\del(y^2)$, only if \rl{822} is satisfied. In this derivation the variations of $H_{AB}$ must also satisfy \rl{821}.

 For the allowed class of Lagrangians,
 \be
 &&\cL_0(y)=\ga\biggl(3R_{ABCD}(y)R^{ABCD}(y)-6 R_{AB}(y)R^{AB}(y)+ R^2(y)\biggr),\nn\\
 \e{823}
 we find the equations ($\ga\neq0$)
 \be
 &&\del(y^2)\biggl(\Box^2H^C_{\;\;C}\eta_{AB}-\Box\dif_C\dif_DH^{CD}\eta_{AB}-\dif_A\dif_B\Box H^C_{\;\;C}-2\dif_A\dif_B\dif_C\dif_DH^{CD}-\nn\\&&-3(\Box^2H_{AB}-\Box\dif_A\dif_CH^C_{\;\;B}-\Box\dif_B\dif_CH^C_{\;\;A})+y_AS_B(y)+y_BS_A(y)\biggr)=0,\nn\\
 \e{824}
 where  $S_A$ are arbitrary unphysical  functions which      enters since the variation of $H_{AB}$ satisfies $y^A\del H_{AB}=0$. Self-consistency of \rl{824} requires that $S_A$  satisfies the weak relations
 \be
 &&(y\cdot\dif+5)S_A(y)=y^2f_A(y), \quad y^AS_A(y)=y^2g(y).
 \e{8241}
 The equations \rl{824} determine $S_A$ in terms of $H_{AB}$ and leave only 15 true equations for $H_{AB}$. (However, the condition $y^AH_{AB}=0$ represents 6 more conditions on $H_{AB}$.)
 One may easily check that \rl{824} is consistent with \rl{821}. We notice also that the equations \rl{824} are invariant under the special gauge transformations \rl{8201}. Only $S_A$ transforms to $S'_A$, but both are equally arbitrary. Gauge invariance follows since the values of $S_A$ may be ignored.

 In appendix C we reduce the action of \rl{823} and the equations \rl{824} to expressions only involving the coordinates $x^{\mu}$ by means of the transformation \rl{04},\rl{05}. The equations \rl{824} become then the equations from linear conformal gravity in $d=4$, and the action reduces to the action for linear conformal gravity in $d=4$, which is a perfectly consistent result.

   \setcounter{equation}{0}
\section{Coordinate independent form of the $s=2$ theory}
When we treated the $s=2$ model in $d=4$ in section 6.2 we found that the Riemann tensor appeared naturally  also in the manifestly six dimensional theory. However, this Riemann tensor was expressed in terms of flat coordinates. In order to be able to describe gravity in arbitrary coordinates and view the Riemann tensor as a general tensor even in six dimensions some of the conditions in section 6.2 have to be generalized. 
The strong transversality condition \rl{0832} has \eg to be generalized as follows
\be
&&y^AR_{ABCD}(y)=0\quad \longrightarrow\quad Y^A(y)R_{ABCD}(y)=0,
\e{901}
where $Y^A(y)$ is a general vector field which for flat coordinates reduces to $y^A$. 
In fact, the homogeneity and transversality of a general tensor field $A_{ABC\cdots}(y)$ have to be generalized according to the following rules
\be
&&(y\cdot\dif-n)A_{ABC\cdots}(y)=0\quad \longrightarrow\quad(Y(y)\cdot D-n)A_{ABC\cdots}(y)=0,\nn\\
&&y^AA_{ABC\cdots}(y)=0\quad\longrightarrow\quad Y^AA_{ABC\cdots}(y)=0,
\e{902}
where $D_A$ is the covariant derivative. In particular the homogeneity condition \rl{314}, \rl{710} generalizes as follows
\be
&&y\cdot\dif G_{AB}(y)=0\quad\longrightarrow\quad Y(y)\cdot DG_{AB}(y)=0,
\e{903}
which is trivially satisfied since
\be
&&D_C G_{AB}(y)=0.
\e{904}
This generalizes the flat coordinate property $\dif_C\eta_{AB}=0$. The condition \rl{901} may easily be solved by the defining properties of the Riemann tensor. In terms of covariant derivatives the condition \rl{901} may be written as
\be
&&(D_{A}D_{B}-D_{B}D_{A})Y_C(y)\equiv-Y^D(y)R_{DCAB}(y)=0.
\e{905}
\rl{904} suggests then the following solution
\be
&&D_{A}Y_{B}(y)=G_{AB}(y),
\e{906}
which also is the direct generalization of the flat coordinate property
\be
&&\dif_Ay_{B}=\eta_{AB}.
\e{907}
Property \rl{906}  implies furthermore
\be
&&0=(D_{A}Y_{B}(y)-D_{B}Y_{A}(y))=(\dif_AY_{B}(y)-\dif_BY_A(y)),
\e{908}
which in turn implies
\be
&&Y_{A}(y)=\half\dif_AU(y),
\e{909}
where $U(y)$ is a scalar field. The factor half is chosen in order to let $U(y)$ generalize $y^2$ in flat coordinates.

A different way to derive the above properties is given below.

 \setcounter{equation}{0}
 \section{Coordinate independent form of the conformal particle in  external fields}
 Consider the conformal particle in external tensor fields up to rank two in a coordinate independent form. The appropriate Lagrangian is then (generalizes only slightly the forms given in \cite{Bars:2000cv} and \cite{Britto-Pacumio:1999ax})
 \be
 &&L(\tau)={1\over 2v} \dy^AG_{AB}(y)\dy^B-\half v g^k\phi^k(y)-eA_B(y)y^B+\la U(y),
 \e{324}
 where $g$ and $e$ are coupling constants. Apart from the insertion of $G_{AB}(y)$ we have replaced  $y^2$ by a general scalar $U(y)$ in order to investigate the possibility of a coordinate invariant formulation. From \rl{324} we find the conjugate momentum
 \be
 &&p_A={1\over v}G_{AB}(y)\dy^B-eA_A(y).
 \e{325}
 As before we require $G_{AB}$ to have the inverse $G^{AB}$. The Hamiltonian becomes then
 \be
 &&H=\half v(p_A+eA_A(y))G^{AB}(y)(p_B+eA_B(y))+\half v g^k\phi^k(y)-\la U(y),\nn\\
 \e{326}
 and in addition we 
find  the constraints (cf.\cite{Britto-Pacumio:1999ax,Bars:2000cv})
 \be
 &&\chi_1\equiv (p_A+eA_A(y))G^{AB}(y)(p_B+eA_B(y))+g^k\phi^k(y),\nn\\
 &&\chi_2\equiv \half (p_A+eA_A(y))G^{AB}(y)\dif_BU(y),\quad \chi_3\equiv U(y).
 \e{327}
 As in section 4 we require $\chi_i$ to satisfy the SL(2,R)-algebra \rl{2021}. If we define the vector field $Y^A(y)$ by (cf \cite{Bars:2000cv})
 \be
 &&Y^A(y)\equiv \half G^{AB}(y)\dif_B U(y),
 \e{328}
 it may be identified with the vector field $Y^A(y)$ used in the previous section. The scalar $U(y)$ used here is then the same scalar  as in  \rl{909}. With \rl{328} in \rl{327}
  the SL(2,R)-algebra of the constraints $\chi_i$ requires
 \be
 &&(Y^A(y)\dif_A+{2/k})\phi(y)=0,
 \e{329}
 \be
 &&Y^A(y)F_{AB}(y)=0,\quad F_{AB}(y)\equiv \dif_AA_B(y)-\dif_BA_A(y),
 \e{330}
 \be
 &&Y^A(y)\dif_AU(y)=2U(y),
 \e{331}
  \be
 &&Y^C(y)\dif_CG^{AB}(y)-\dif_C Y^A(y)G^{CB}(y)-\dif_CY^B(y)G^{CA}(y)=-2G^{AB}(y),\nn\\
 \e{332}
 Condition \rl{329} is obviously the coordinate invariant generalization of the homogeneity condition \rl{303}, and \rl{330} is then the corresponding generalization of \rl{307}. Notice that \rl{330} may be solved by the conditions ($D_A$ is the covariant derivative)
 \be
 &&Y^A(y)A_A(y)=0,\quad (Y^A(y)D_A+1)A_B(y)=0,
 \e{333}
 which then generalizes \rl{308}. The last condition \rl{332} is easily solved, if we  impose the conditions
 \be
 &&D_AY^B(y)=\del_A^B \quad \Leftrightarrow\quad D_A\dif_BU(y)=2G_{AB}(y),
 \e{334}
 which is equivalent to \rl{906} and \rl{909}.
 (The last condition generalizes $\dif_A\dif_By^2=2\eta_{AB}$, and is also consistent with the homogeneity condition \rl{331}.) First we notice that $\dif_C$ may be replaced by the covariant derivative $D_C$ in \rl{332}. Then, since we trivially have
 \be
 &&Y^C(y)D_CG^{AB}(y)=0,
 \e{335}
  \rl{332} follows from \rl{334}.

 \setcounter{equation}{0}
\section{Actions for general tensor fields in curved dynamical backgrounds and manifestly conformal gravity}
The results of the previous two sections suggest that there also is a coordinate invariant action principle
in $d+2$ dimensions. The actions should then have the form
\be
&&A=\int dy\sqrt{-G}\del(U(y))\cL(y),
\e{1101}
where
\be
&&G\equiv \det G_{AB}(y).
\e{1102}
$U(y)$ is the general scalar introduced in the previous sections. $\cL(y)$ is required to be a general scalar expressed in terms of general tensor fields $A_{ABC\cdots}(y)$ which are subject to the coordinate invariant conditions
\be
&&(Y\cdot D-n)A_{ABC\cdots}(y) =0,\quad Y^AA_{ABC\cdots}(y)=0,\nn\\
&&Y^BA_{ABC\cdots}(y)=0,\quad Y^CA_{ABC\cdots}(y)=0, \quad {\rm etc}.,
\e{1103}
where the constant $n$ is the generalized degree of homogeneity. $Y^A(y)$ is the general vector field introduced in the previous sections.  \rl{1103}
 generalizes \rl{8021}. The coordinate invariant hypercone $U(y)=0$ may also be written as \newline $Y^A(y)Y_A(y)=0$ since
 \be
 &&U(y)=Y^A(y)Y_A(y)
 \e{1104} 
 due to \rl{331}. Notice the homogeneities
 \be
 &&Y\cdot DY^A(y)=Y^A(y),\quad Y\cdot D U(y)=2U(y),\nn\\&& Y\cdot D G_{AB}(y)=0.
 \e{1105}
 Since these relations imply
 \be
 &&Y\cdot D G=0,\quad Y\cdot D\del(U(y))=-2\del(U(y))
 \e{1106}
 the condition \rl{802} generalizes to
 \be
 &&(Y\cdot D+d)\cL(y)=0.
 \e{1107}
 Again we have to require that all fields in the action \rl{1101} is possible to restrict to the generalized hypercone $U=0$. This is possible if we require the action to be invariant under transformations of the form (special gauge transformations off the generalized hypercone)
 \be
&&A_{ABC\cdots}(y) \;\longrightarrow\;A'_{ABC\cdots}(y) = A_{ABC\cdots}(y) +U(y)\tilde{A}_{ABC\cdots}(y),\nn\\
&&(Y\cdot D-n+2)\tilde{A}_{ABC\cdots}(y) =0,\quad Y^A\tilde{A}_{ABC\cdots}(y)=0,\nn\\
&& Y^B\tilde{A}_{ABC\cdots}(y)=0,\quad Y^C\tilde{A}_{ABC\cdots}(y)=0, \quad {\rm etc}.,
\e{11071}
for all fields involved. (This invariance is  valid for the external fields in section 10.) We expect that this also implies that the equations are strictly valid on the generalized hypercone $U=0$. Notice that
 in deriving equations from the action \rl{1101} we have to use variations which are consistent with the imposed conditions \rl{1103}.

 \subsection{Manifestly conformal gravity}
 In all  variations of the involved fields $Y^A(y)$ and $U(y)$ are kept fixed. However, this is not automatic
 when we vary the metric tensor $G_{AB}(y)$. Here  we have to use restricted variations to secure that they remain fixed since
 \be
 &&Y^AG_{AB}(y)=Y_B(y)
 \e{1113}
is valid by definition. For the variations of $G_{AB}$ we propose the following conditions
 \be
 &&Y^A\del G_{AB}(y)=Y_A(y)\del G^{AB}(y)=0, \quad Y\cdot D\del G_{AB}(y)=0,
 \e{1114}
 which are of the general type \rl{1103}. The first conditions are the same as used for linear gravity in section 8, but now applied to a general background. Assuming \eg $Y_A$ to be fixed
the first conditions imply that $Y^A$ is fixed and that
 \be
 &&\del U(y)=\del (Y^A(y)Y_A(y))=Y_{A}(y)Y_B(y)\del G^{AB}(y)=0.
 \e{1115}
 The generalized hypercone is therefore kept fixed by variations satisfying \rl{1114}.
 
 Since $D_A$ has the homogeneity minus one it is easily derived from the defining properties that the Riemann tensor satisfies the general homogeneity (cf \rl{711})
  \be
 &&(Y\cdot D+2)R_{ABCD}(y)=0.
 \e{1116}
 We find therefore that the general ansatz \rl{817}, \ie
 \be
 &&\cL(y)=\al R_{ABCD}(y)R^{ABCD}(y)+\beta R_{AB}(y)R^{AB}(y)+\ga R^2(y),\nn\\
 \e{1117}
 satisfies the condition \rl{1107} for $d=4$.
 
  It remains to investigate the invariance under the generalized special gauge transformations \rl{11071} which for the metric tensor $G_{AB}$ takes the form
 \be
&&G_{AB}(y)\;\;\ra\;\;G'_{AB}(y)=G_{AB}(y)+U(y)\tilde{H}_{AB}(y),\nn\\
\e{1118}
where the arbitrary tensor $\tilde{H}_{AB}$  satisfies
\be
&&Y^A(y)\tilde{H}_{AB}(y)=0,\quad (Y\cdot D+2)\tilde{H}_{AB}(y)=0,
\e{1119}
which secures \rl{1113}. The Riemann and Ricci tensors as well as the curvature scalar transform under \rl{1118} in arbitrary dimensions $d$ as follows: (ignoring $U$-terms)
\be
&&R_{ABCD}(y)\;\;\ra\;\;R_{ABCD}(y)+G_{AC}(y)\tilde{H}_{BD}(y)+G_{BD}(y)\tilde{H}_{AC}(y)-\nn\\&&-G_{BC}(y)\tilde{H}_{AD}(y)-G_{AD}(y)\tilde{H}_{BC}(y)+\nn\\&&+Y_A(y)(D_C\tilde{H}_{BD}(y)-D_D\tilde{H}_{BC}(y))+\nn\\&&+Y_B(y)(D_D\tilde{H}_{AC}(y)-D_C\tilde{H}_{AD}(y))+\nn\\&&+Y_C(y)(D_A\tilde{H}_{BD}(y)-D_B\tilde{H}_{AD}(y))+\nn\\&&+Y_D(y)(D_B\tilde{H}_{AC}(y)-D_A\tilde{H}_{BC}(y))+\nn\\&&+Y_A(y)Y_D(y)\tilde{H}_{\;\;C}^{M}(y)\tilde{H}_{MB}(y)+Y_B(y)Y_C(y)\tilde{H}_{\;\;A}^M(y)\tilde{H}_{MD}(y)-\nn\\&&-Y_A(y)Y_C(y)\tilde{H}_{\;\;D}^M(y)\tilde{H}_{MB}(y)-Y_B(y)Y_D(y)\tilde{H}_{\;\;A}^M(y)\tilde{H}_{MC}(y),
\e{11191}
which implies
\be
&&R_{BD}(y)\;\;\ra\;\;R_{BD}(y)+(d-2)\tilde{H}_{BD}(y)+G_{BD}(y)\tilde{H}^C_{\;\;C}(y)+\nn\\&&+Y_B(y)D_D\tilde{H}^C_{\;\;C}(y)+Y_D(y)D_B\tilde{H}^C_{\;\;C}(y)-\nn\\&&-Y_B(y)D_C\tilde{H}^C_{\;\;D}(y)-Y_D(y)D_C\tilde{H}^C_{\;\;B}(y)-\nn\\&&-Y_B(y)Y_D(y)\tilde{H}_{AC}(y)\tilde{H}^{AC}(y),
\e{11192}
and
\be
&&R(y)\;\;\ra\;\;R(y)+2(d-1)\tilde{H}^C_{\;\;C}(y).
\e{11193}
These results imply (ignoring $U$-terms)
\be
&&R^{ABCD}(y)R_{ABCD}(y)\;\;\ra\;\;R^{ABCD}(y)R_{ABCD}(y)+8R_{AB}(y)\tilde{H}^{AB}(y)+\nn\\&&+4(d-2)\tilde{H}_{AB}(y)\tilde{H}^{AB}(y)+4(\tilde{H}^C_{\;\;C}(y))^2,
\e{1120}
\be
&&R_{BD}(y)R^{BD}(y)\;\;\ra\;\;R_{BD}(y)R^{BD}(y)+2(d-2)R_{AC}(y)\tilde{H}^{AC}(y)+\nn\\&&+2R(y)\tilde{H}^C_{\;\;C}(y)+(d-2)^2\tilde{H}_{AB}(y)\tilde{H}^{AB}(y)+(3d-4)(\tilde{H}^C_{\;\;C}(y))^2,\nn\\
\e{1121}
\be
&&R^2(y)\;\;\ra\;\;R^2(y)+4(d-1)R(y)\tilde{H}^C_{\;\;C}(y)+4(d-1)^2(\tilde{H}^C_{\;\;C}(y))^2.\nn\\
\e{1122}
  Inserting these expressions into the Lagrangian \rl{1117} we find that $\cL(y)$ is invariant for $d=4$ and
\be
&&\al=3\ga,\quad \beta=-6\ga,
\e{1123}
which agrees with the linear gravity result \rl{822}. In fact, in the linear case we have exactly the same expressions as \rl{1121}-\rl{1123} except that indices are raised, lowered and contracted by means of the flat metric $\eta_{AB}$.  We expect therefore that the action corresponding to \rl{1117} for the values \rl{1123}  yield equations on the general hypercone $U(y)=0$. Thus, the Lagrangian \rl{823} should be a satisfactory Lagrangian for nonlinear tensors and general metrics $G_{AB}(y)$ provided the variations in the corresponding action \rl{1101} are restricted according to \rl{1114}.

\subsection{Previous conformal field theories in a curved dynamical background}
 All models considered in section 8 may  be generalized to a curved dynamical background. Complications occur mainly for the scalar field theory. The free scalar theory  \rl{804} generalizes   to
 \be
 &&\cL_0(y)=\half \phi(y)D^A\dif_A\phi(y),
 \e{1108}
 where $\phi(y)$ now is a general scalar. A variation of $\phi(y)$ subject to
 \be
 &&(Y\cdot D+d/2-1)\phi(y)=0,
 \e{1109}
 which generalizes \rl{803}, reproduces
 \be
 &&D^A\dif_A\phi(y)=0
 \e{1110}
 on $U(y)=0$, which in turn generalizes \rl{805}. Both the action integral  to \rl{1108} and \rl{1110} are invariant under
 \be
 &&\phi(y)\;\;\ra\;\;\phi(y)+U(y)\tilde{\phi}(y).
 \e{1124}
 However, neither  \rl{1108} nor \rl{1110} are invariant under \rl{1118}. In fact, for \rl{1108} we find
 \be
 &&\cL_0(y)\;\;\ra\;\;\cL_0(y)+\half \biggl(1-{d\over 2}\biggr)\tilde{H}^C_{\;\;C}(y)\phi^2(y).
 \e{1125}
 On the other hand, if we replace \rl{1108} by
 \be
 &&\cL(y)=\half \phi(y)D^A\dif_A\phi(y)+\ga R(y)\phi^2(y),
 \e{1126}
 where $\ga$ is a constant, we find from \rl{1125} and the formula \rl{11193} that $\cL(y)$ and its equation of motion are invariant for
 \be
 &&\ga={d-2\over 8(d-1)}
 \e{1127}
 in agreement with the well-known result in $d$ dimensions.
 
 In the case of spinors like the $s=1/2$ field considered in section 8.2 we need vielbein fields.
 The vielbein $V^a_A(y)$ satisfies the properties
 \be
 &&\eta_{ab}V^a_A(y)V^b_B(y)=G_{AB}(y), \quad D_AV^b_B(y)=0,
 \e{1111}
 where we use small letters for flat indices and capitals for curved indices. The Lagrangian \rl{808} \eg generalizes then to ($\al$, $\beta$ and $\ga$ are spinor indices)
 \be
 &&\cL_0(y)=\half \bar{\psi}_{\al}(y)\eta_{ac} Y^A(y)V_A^a(y)\Gamma^c_{\al\beta}\eta_{bd}V^b_C(y)g^{CB}(y)D_B\Gamma^d_{\beta\ga}\psi_{\ga}(y).\nn\\
 \e{1112}

 \setcounter{equation}{0}
\section{Conclusions}
In this paper we have considerably extended the manifestly conformal theory both what concerns spinning particles and field theory. Let us specify what is new and some remaining questions:

The Lagrangians for the scalar particle and the supersymmetric one in section 3 were simplified and considered for arbitrary dimensions as compared to the original paper \cite{Marnelius:1979conformal}. The treatment of external fields in section 4  for symmetric tensor fields of rank two and higher was different from \cite{Bars:2001um} but with consistent results. The treatment of spinning conformal particles in section 5 was improved and considered for arbitrary dimensions as compared to the original treatments \cite{Marnelius:1990de,Martensson:1992ax}. Its Dirac quantization has not been considered before, and the derivation of conformal tensor fields in section 6 is also new. Previously a BRST quantization was performed in \cite{Martensson:1992ax}. (In appendix D we treated a related relativistic particle model for arbitrary spins, which we quantized  in a simple way and formally generalized the results to arbitrary spacetime dimensions. Previously it was quantized using BRST in \cite{Marnelius:1988ab}.)  The clash between the properties of external tensor fields and the fields produced in section 6 made us clarify the relation between homogeneity and the order of equations in section 7. It turned out that symmetric tensor fields of order $s(>0)$ requires equations of order $2s$. The basic assumption made in previous sections that all spinning particle models lead to second order equations seem therefore to be wrong. We have not tried to find particle models leading to higher order equations and this possibility is therefore open. For spin two ($s=2$ in $d=4$) we took all symmetry properties from the particle model excluding conditions connected to the second order equation like homogeneity and Lorentz like conditions. Eventually we arrived at the Riemann tensor in six dimensions ($d=4$). We were then ready to investigate the field theory case. The specifications of the properties of manifestly conformal fields given in section 8 are much more precise compared to what one previously has considered. The special gauge transformations off the hypercone have never been used to such an extent before. They turned out to be crucial for the second rank tensor theory.    (Actually, second rank tensors have hardly been treated at all in this context before.) The precise action principle in section 9 is new although two of the actions were given in \cite{Marnelius:1980}. The previously proposed actions in \cite{Budini:1979} agree apart from a delta function in the measure what concerns scalar and vector fields, but the treatments of spinor  fields differ. (The reduction to four dimensional spacetime in  \cite{Budini:1979} only agrees for scalar fields.) Our treatment of vector and tensor fields in the actions and the corresponding equations of motion are new and crucial for  consistent results, as well as our reductions in appendices B and C. Our treatment of coordinate invariant theories within the manifestly conformal framework is new (sections 10 and 12). (The conditions for external fields in section 11 are not new but our way to solve them are.) Our proposal for a manifestly conformal gravity is new (cf \cite{Preitschopf:1998}).
We have not investigated its reduction to four dimensional spacetime which therefore remains. However, for  its linearized form we have essentially given all details. 

Manifestly conformal field theory is open for further investigations along the lines of the present paper.

 \begin{appendix}
\newpage
%\small
\section{The reduction  of manifestly conformal fields in $d+2$ dimensions to $d$ dimensional spacetime}
In this appendix we give the main ingredients involved when reducing a $d+2$ dimensional manifestly conformal field on the conformal space to a conformal field on a $d$-dimensional spacetime. We follow then mainly \cite{Mack:1969}. We also comment on the reduction of the action.

Consider a general field $F^i(y)$ defined on the conformal space. The superscript $i$ denotes tensor or/and spinor indices, depending on the type of field. Note that $i$ should be consistent with a representation of the conformal group SO($d$,2). First we always require the field to be a homogeneous function of the coordinates $y^A$ in the sense 
\be
&&y^A \frac{\partial}{\partial y^A} F^i(y) = n F^i(y),
\e{homo_y}
where $n$ denotes the degree of homogeneity.
From the relations \rl{04} and \rl{05} with $R=1$ we find 
\be
&&y^A \frac{\partial}{\partial y^A} = \ga \frac{\partial}{\partial \ga} + 2 \rho \frac{\partial}{\partial \rho},\quad \ga\equiv y^-,\quad \rho\equiv y^2.
\e{hom}
As a first consideration let us assume that we may choose a gauge in which 
 the field $F^i(y)$ does not depend on $y^2$, \ie
\be
&&\left.F^i(y)=F^i(y)\right |_{\rho=0},\quad \left.\dif_{\rho}F^i(y)\right |_{\rho=0}=0,\nn\\
&&\left.\dif^2_{\rho}F^i(y)\right |_{\rho=0}=0, \;\;{\rm etc.},\quad \rho\equiv y^2.
\e{concond}
The required special gauge transformation \rl{0803}/\rl{80211} does in principle allow for this possibility.
The condition \rl{concond} implies  together with \rl{hom}
 that the homogeneity relation~\rl{homo_y} may be rewritten as
\be
&&\ga \frac{\partial}{\partial \ga} F^i(y) = n F^i(y).
\e{homo_ga}
 This relation allows us finally to define a field $\tilde{f}^i(x)$ which only depends on the $d$-dimensional coordinates $x^{\mu}$ as follows
\be
&&\tilde{f}^i(x) \equiv \ga^{-n} F^i(y).
\e{phi_tilde}
Now when $i$ contains tensor indices we can no longer consistently impose \rl{concond} due to the strong transversality conditions. Here, as explained in section 8, we must split the field  in several terms involving fields and explicit coordinates $y^A$, and  then impose \rl{concond} on all these component  fields separately leaving the $y^2$ dependence in the explicit coordinates $y^A$. Notice that this splitting does not change the degrees of freedom. It is just a way to specify how to get down to $d$ dimensions.
Now, since all these fields satisfy strong homogeneity conditions we find \rl{homo_ga} for each of them (with different $n$). For each of these fields we may then define $\tilde{f}^i(x)$ -fields depending on $x^{\mu}$ according to \rl{phi_tilde}. However,  the original field does not depend entirely on $x^{\mu}$ when we follow this procedure.

For non-scalar fields  \rl{phi_tilde} is not enough to produce a consistent spacetime field. The reason for this is the following: The operator that generates translations, $P_{\mu}$, which is equal to $J_{+ \mu}$, is supposed to act only differentially on a spacetime field. Now this latter operator may be divided into two parts according to
\be
&&J_{+ \mu} = L_{+ \mu} + S_{+ \mu}.
\e{translation}
In this expression, $L_{+ \mu}$ is the differential (orbital) piece while $S_{+ \mu}$ is the intrinsic (spin) piece. The intrinsic piece is non-differential and should, therefore, be removed. This may be done by  defining the true field on the $d$-dimensional spacetime as follows:
\be
&&f^i(x) \equiv V(x) \tilde{f}^i(x) = \ga^{-n} V(x) F^i(y),
\e{fields}
where the operator $V(x)$ is defined by
\be
&&V(x) \equiv \exp(- i x^\mu S_{+ \mu}).
\e{V}
The field $f^i(x)$ behaves then as expected under translations, and depends only on the spacetime coordinates $x^\mu$.

Finally, it should be mentioned that the projected field $f^i(x)$ in \rl{fields} in general will yield unphysical components in addition to the expected physical ones.
These are present \eg if $i$ is a vector index $A$, in which case we get two additional components ($F^+$ and $F^-$) compared to a vector field in $d$ dimensions (see spin one and two below). Following the prescription of~\cite{Mack:1969}, the unphysical components may be  projected out by the additional condition
\be
&&(S_{-\mu} f)^i = 0
\e{unphysical}
for all physical values of $i$ and all values of $\mu$. However,  although we do not explicitly impose   such additional "gauge fixing" conditions        in this paper, the final formulation will always allow for them.

The actions may be reduced to $d$ dimensions as follows: Choose the Lagrangian density as
\be
&&\cL(y)=\cL(y,y^2=0).
\e{a010}
From
\be
&&(y\cdot\dif+d)\cL(y)=0,
\e{a011}
we find 
\be
&&\cL(x)=\ga^{-d}\cL(y,y^2=0)
\e{a012}
following the steps \rl{homo_ga} and \rl{phi_tilde}. Since the Lagrangian is a scalar there are no further modifications. The action becomes
\be
&&A=\int d y \del(y^2)\cL(y)=\int d y \del(y^2)\cL(y,y^2=0)=C\int dx \cL(x),
\e{a013}
where
\be
&&C=\half\int {d \ga\over \ga}=\half \int {d x^5\over x^5}.
\e{a014}
This is a logarithmically divergent constant. Notice, however, that the point transformation $y^A\;\ra\;x^A$ in section 2 is singular at $x^5=0$. Thus, $x^5$ should be positive or negative. In \cite{Budini:1979} an extra delta function that fixes $x^5$ was inserted. This is not completely unnatural. The conformal particle models always allow for a gauge choice of $x^5$. For instance, $x^5=f(x)$ yields fields in a curved background with the metric $g_{\mu\nu}(x)=f^2(x)\eta_{\mu\nu}$ (see section 2). In \cite{Marnelius:1980}  it was shown how the fields may be reduced to  fields in such a space. Here, it could be natural with a delta function $\del(x^5-f(x))$, but then as a part of the reduction procedure. Although, \rl{a013} and \rl{a014} yield a field theory in a compactified Minkowski space \cite{Penrose:1965am}, we may always change the background by a conformal transformation.

 \setcounter{equation}{0}
\section{The reduction  of manifestly conformal spin one fields to $d=4$ dimensions}

As a simple but non-trivial example of the reduction method described above, we will now consider a vector field $a_{\mu}(x)$ in four dimensions. The manifestly conformal field corresponding to this field is denoted as $A_{A}(y)$, with the corresponding field strength given by (cf.~\rl{307})
\be
&& F_{AB} = \frac{\partial}{\partial y^A} A_B - \frac{\partial}{\partial y^B} A_A.
\e{F_AB}
For $A_B$ we require the degree of homogeneity to be $n=-1$ and impose the strong transversality condition
\be
&& y^A A_A(y) = 0,
\e{ydotA}
as in \rl{308}.  Now, the strong condition \rl{ydotA} implies that $A_A$ necessarily has a $y^2$ dependence. However, as explained in section 8, if we write $A_A$ as follows
\be
&&A_A(y)=V_A(y)-y_A\phi(y),\nn\\
&&(y\cdot\dif+1)V_A(y)=0,\quad (y\cdot\dif+2)\phi(y)=0,
\e{appB}
then $V_A(y)$ and $\phi(y)$ may be chosen to be independent of $y^2$. (Notice that this $V_A$ satisfies the corresponding weak condition to \rl{ydotA} (see \rl{aminus}).) To $V_A$ we may then apply the steps \rl{concond}-\rl{V}. In order to apply \rl{fields} we need the action of 
the intrinsic SO(4,2) generator $S_{AB}$ on a vector field (see \rl{503}). It is given by
\be
&& S_{AB} A_C = -i \left( \eta_{CA} A_B - \eta_{CB} A_A \right).
\e{SonA}
($S_{AB}$ satisfies the SO(4,2) algebra in \rl{5041}.)
This implies then
\be
V_{\mu}(y,y^2=0) & = & \frac{1}{\ga} \left( a_\mu(x) - x_\mu a_+ (x) \right), \nn \\
V_+(y,y^2=0) & = & \frac{1}{\ga} a_+(x) , \nn \\
V_-(y,y^2=0) & = & \frac{1}{\ga} \left( a_- (x) - x^\mu a_\mu(x)  + \frac{1}{2} x^2 a_+(x) \right),
\e{Aa}
where $a_{\mu}$, $a_+$, and $a_-$ only depend on $x^{\mu}$. 
It is straightforward to find the inverse relations. Condition \rl{ydotA} implies furthermore
\be
&&y^AV_A(y)=y^2\phi(y),\nn\\
&& a_-(x) =0, \quad \phi(y,y^2=0)=-{1\over 2\ga^2}a_+(x) ,
\e{aminus}
by means of \rl{Aa}.

The natural Lagrangian in the conformal space is given in \rl{810}, which for $d=4$ becomes
\be
&&\cL(y)={1\over4}F_{AB}(y)F^{AB}(y).
\e{Lag}
Inserting \rl{appB} into $F_{AB}$ we find
\be
&&F_{AB}(y)=W_{AB}(y)+L_{AB}\phi(y),\nn\\
&&W_{AB}(y)=\dif_AV_B(y)-\dif_BV_A(y),\quad L_{AB}\phi(y)=(y_A\dif_B-y_B\dif_A)\phi(y).\nn\\
\e{b1}
This implies that the Lagrangian \rl{Lag} may be written as
\be
&&\cL(y)={1\over4}W_{AB}(y)W^{AB}(y)+2\phi^2(y).
\e{Lag2}
By means of \rl{Aa},  $W_{AB}W^{AB}$ reduces  to
\be
&&W_{AB}(y)W^{AB}(y)={1\over \ga^4}(f_{\mu\nu}(x)f^{\mu\nu}(x)-2a_+^2(x)),\nn\\
&& f_{\mu\nu}(x)\equiv \dif_{\mu}a_{\nu}(x)-\dif_{\nu}a_{\mu}(x).
\e{b2}
Hence, \rl{Lag2} with  \rl{aminus}  implies the reduction (see \rl{a012})
\be
&&\cL(y)={1\over4}F_{AB}(y)F^{AB}(y)\quad\ra\quad {1\over4\ga^4}f_{\mu\nu}(x)f^{\mu\nu}(x),
\e{b3}
which according to \rl{a013} implies that the original action reduces to
\be
&&A=C\int d^4x{1\over 4}f_{\mu\nu}(x)f^{\mu\nu}(x),\quad f_{\mu\nu}(x)=\dif_{\mu}a_{\nu}(x)-\dif_{\nu}a_{\mu}(x).
\e{b4}
This reduction is different from previous attemts \cite{Budini:1979} in which the Lagrangian density was identified with
\be
&&\cL(y)={1\over4}W_{AB}(y)W^{AB}(y)\quad\ra\quad {1\over4\ga^4}(f_{\mu\nu}(x)f^{\mu\nu}(x)-2a_+^2(x)),\nn\\
\e{b5}
where the auxiliary field $a_+$ enters, which it does not in \rl{b4}. Only in \rl{b4} are we free to impose $a_+=0$ as required by the condition \rl{unphysical}.

The equations of motion from \rl{Lag} are on the hypercone, and are given by
\be
&&\dif^AF_{AB}(y)=y_BS(y),
\e{eq}
where $S$ is arbitrary apart from the fact that self-consistency requires the weak homogeneity
\be
&&(y\cdot\dif+4)S(y)=y^2f(y). 
\e{b51}
$S$ is an unphysical  function that would be zero if all components of $A_A$ were independent. However, the transversality condition \rl{ydotA} makes one component dependent. The tensor equations \rl{eq} with $S$ follow then   since the variation of $A_A$ also satisfies \rl{ydotA}.
Only  $S$ is affected by the special gauge transformations off the hypercone
\be
&&A_A(y)\quad\ra\quad A'_A(y)=A_A(y)+y^2\tilde{A}_A(y),\nn\\
&&(y\cdot\dif+3)\tilde{A}_A(y)=0,\quad y^A\tilde{A}_A(y)=0.
\e{b6}
On \rl{eq} they yield on the hypercone ($S$ and $S'$ are equally arbitrary)
\be
&&\dif^AF_{AB}(y)=y_BS(y)\quad\ra\quad \dif^AF_{AB}(y)=y_BS'(y),\nn\\
&&S'(y)=S(y)-2\dif^A\tilde{A}_A(y).
\e{b7}
$S$ reduces the number of equations for $A_B$ and the gauge invariance means that its value may be ignored  at the end.

To reduce the equations \rl{eq} to $d=4$ we first  apply the decomposition \rl{appB} to \rl{eq} which yields
\be
&&\dif^AW_{AB}(y)+2\dif_B\phi(y)=y_BS''(y), \quad S''(y)=S(y)+\Box\phi(y).
\e{eq2}
Then using \rl{Aa} and \rl{aminus} we find finally that \rl{eq} leads to the reduced equations
\be
&&\dif^{\mu}f_{\mu\nu}(x)=0,
\e{b8}
which is consistent with the result \rl{b4} for the reduction of the action. In the process the arbitrary function $S$ becomes
\be
&&S''=-{1\over\ga^4}\Box_x a_+(x) \quad\Leftrightarrow\quad S=-{1\over 2\ga^4}\Box_xa_+(x),
\e{b9}
which are purely unphysical expressions that should be ignored.

Finally, both \rl{unphysical} and \rl{b4}/\rl{b8}/\rl{b9} tell us that $a_+$ is unphysical. The transversality condition \rl{ydotA} makes  $a_-$ unphysical according to \rl{aminus}.

 \setcounter{equation}{0}
\section{The reduction  of  linear manifestly conformal gravity  to $d=4$ dimensions}

After exploring spin one, we now apply these ideas to linearized manifestly conformal gravity in $d=4$. The field we want to get is the symmetric rank two tensor $h_{\mu\nu}(x)$, while the corresponding manifestly conformal field from which we start is $H_{AB}(y)$, which also is symmetric. We impose the conditions of homogeneity of degree zero and strong transversality, \ie
\be
&&y\cdot\dif H_{AB}(y)=0,\quad y^A H_{AB} = 0.
\e{ydotH}
These conditions restrict the basic special gauge transformations to
\be
&&H_{AB}(y)\quad\ra\quad H_{AB}(y)+y^2\tilde{H}_{AB}(y),\nn\\
&&(y\cdot\dif+2)\tilde{H}_{AB}(y)=0,\quad y^A\tilde{H}_{AB}(y)=0.
\e{c01}
The last property implies that we cannot choose the gauge \rl{concond}. Instead we have to do as in the spin one case: we have to split $H_{AB}$. The appropriate splitting is (see section 8)
\be
&&H_{AB}(y)=K_{AB}(y)-y_AV_B(y)-y_BV_A(y)+y_Ay_B\phi(y),\nn\\
&&(y\cdot\dif+1)V_A(y)=0,\quad (y\cdot\dif+2)\phi(y)=0.
\e{c02}
The transversality condition in \rl{ydotH} is then solved by
\be
&&y^AK_{AB}(y)=y^2V_B(y),\quad y^BV_B(y)=y^2\phi(y).
\e{c03}
The gauge invariance under \rl{c01} allows us then to fix $K_{AB}$, $V_B$, and $\phi$ according to \rl{concond}, and for them we may follow the steps \rl{homo_ga}-\rl{phi_tilde}. In order to apply the step \rl{fields} we need the 
 action of the intrinsic SO(4,2) generator $S_{AB}$ on a symmetric rank two tensor. It is
\be
&& S_{AB} H_{CD} = -i \left( \eta_{CA} H_{BD} - \eta_{CB} H_{AD} + \eta_{DA} H_{BC} - \eta_{DB} H_{AC} \right).
\e{spin2}
Using this expression and \rl{fields} we find  the relations
\be
K_{\mu\nu}(y,y^2=0)& = & h_{\mu\nu}(x) - x_{\mu} h_{\nu +}(x)  - x_\nu h_{\mu+}(x)  + x_{\mu} x_{\nu} h_{++}(x),  \nn\\
K_{\mu+}(y,y^2=0) & = & h_{\mu+}(x)  - x_\mu h_{++}(x),  \nn\\
K_{\mu-}(y,y^2=0) & = & h_{\mu-}(x)  - x^\rho h_{\rho \mu}(x)  -x_{\mu} h_{+-}(x)  + \frac{1}{2} x^2 h_{\mu+}(x)  + {} \nn \\&&{}+ x_\mu x^\rho h_{\rho+}(x)  -
\frac{1}{2} x_\mu x^2 h_{++}(x),  \nn\\
K_{++}(y,y^2=0) & = & h_{++}(x),  \nn\\
K_{+-}(y,y^2=0) & = & h_{+-}(x)  - x^\rho h_{\rho+}(x) + \frac{1}{2} x^2h_{++}(x), \nn \\
K_{--}(y,y^2=0) & = & h_{--}(x)  - 2 x^\mu h_{\mu-}(x)  + x^\mu x^\nu h_{\mu\nu}(x)  + x^2 h_{+-}(x)  -  {} \nn \\&&{}-x^2x^\rho h_{\rho+}(x)  +
\frac{1}{4} (x^2)^2 h_{++}(x).
\e{Hh}

Furthermore, by imposing the transversality condition in \rl{ydotH}, we find that
\be
&& h_{\mu-}(x)  = h_{+-}(x)  = h_{--}(x)  = 0,\quad
V_B(y,y^2=0)=-{1\over 2\ga}K_{+B}(y,y^2=0),\nn\\&& \phi(y,y^2=0)=-{1\over 2\ga}V_+(y,y^2=0)={1\over 4\ga^2}K_{++}(y,y^2=0),
\e{ydotH2}
where we have used \rl{c03}. 
This should be compared with \rl{aminus}.

We proceed by reducing the Lagrangian \rl{823} normalized by $\ga=1$ (this $\ga$ is a constant and not the projective coordinate $\ga=y^-$ above). If we just replace $H_{AB}$ by $K_{AB}$ we find at $y^2=0$
\be
\mathcal{L} & = & \frac{1}{\ga^4} \bigg( 3r_{\mu\nu\rho\sigma}(x)r^{\mu\nu\rho\sigma}(x)-6 r_{\mu\nu}(x) r^{\mu\nu}(x)+ r^2(x) + 9 h_{+\mu}(x)  \square_x h_+^{\phantom{+}\mu}(x)  + {} \nn \\
&& \quad {} + 12 h_{++}(x)  \partial^\rho h_{+\rho}(x)  - 2 r(x)  \partial^\rho h_{+\rho}(x)  - 2 \partial^\rho h_{+\rho}(x)  \partial^\sigma h_{+\sigma}(x)  \bigg),\nn\\
\e{conf_l}
where the quantities $r_{\mu\nu\rho\sigma}(x)$, $r_{\mu\nu}(x)$ and $r(x)$ are the four dimensional Riemann, Ricci and curvature tensors defined in terms of $h_{\mu\nu}$ analogously to the expressions \rl{819} and \rl{820}. However, what we have to do is first to insert the expression \rl{c02} which yields on $y^2=0$
\be
&&\cL(y)=3R_{ABCD}(y)R^{ABCD}(y)-6R_{AB}(y)R^{AB}(y)+R^2(y)-\nn\\&&-24R^{AB}(y)\dif_AV_B(y)+8R(y)\dif_AV^A(y)+28(\dif_AV^A(y))^2+\nn\\&&+12\dif^AV^B(y)\dif_AV_B(y)-60\dif^AV^B(y)\dif_BV_A(y),
\e{c031}
where now the first three terms is the original Lagrangian with $H_{AB}$ replaced by $K_{AB}$. A reduction of the remaining terms together with \rl{conf_l} yields then (see \rl{a012})
\be
\mathcal{L}(y,y^2=0) & = & \frac{1}{\ga^4} \bigg( 3r_{\mu\nu\rho\sigma}(x)r^{\mu\nu\rho\sigma}(x)-6 r_{\mu\nu}(x) r^{\mu\nu}(x)+ r^2(x) \bigg),\nn\\
\e{c04}
which through \rl{a012} and \rl{a013} leads to the linearized conformal gravity in $d=4$. $h_{+\mu}$ does not enter this Lagrangian. According to condition \rl{unphysical} $h_{+\mu}$ is unphysical and should be set to zero. This is consistent with our results.

We should also reduce the equations of motion for  linearized manifestly conformal gravity given in \rl{824}. They are on $y^2=0$
\be
&& \eta_{AB} \left( \Box^2 H^C_{\phantom{C}C} - \Box \dif_C\dif_D H^{CD} \right) -\dif_A\dif_B\Box H^C_{\phantom{C}C}-2\dif_A\dif_B\dif_C\dif_DH^{CD}- {} \nn \\
&& {} - 3\left(\Box^2H_{AB}-\Box\dif_A\dif_CH^C_{\;\;B}-\Box\dif_B\dif_CH^C_{\;\;A} \right) + y_A S_B + y_B S_A = 0, \nn \\
\e{Heom}
where $S_A$ are arbitrary apart from satisfying the weak relations
\be
&&(y\cdot\dif+5)S_A(y)=y^2f_A(y),\quad y^AS_A(y)=y^2g(y),
\e{Sprop}
which are required by consistency.
The functions $S_A$ are unphysical  functions that are determined by \rl{Heom} and therefore reduce the number of equations for $H_{AB}$. They appear due to the restricted variation of $H_{AB}$. These  equations are consistent with \rl{ydotH}, and 
only in the presence of $S_A$ are they invariant under the special gauge transformations \rl{c01} (see subsection 9.4). Although $S_A$ transforms to a new $S_A$ both are equally arbitrary. The equations are gauge invariant since they are  independent of the values of $S_A$.

In order to reduce \rl{Heom} to $d=4$ we have first to insert the equality \rl{c02}. We find then 
\be
&& \eta_{AB} \left( \Box^2 K^C_{\phantom{C}C} - \Box \dif_C\dif_D K^{CD} \right) -\dif_A\dif_B\Box K^C_{\phantom{C}C}-2\dif_A\dif_B\dif_C\dif_DK^{CD}- {} \nn \\
&& {} - 3\left(\Box^2K_{AB}-\Box\dif_A\dif_CK^C_{\;\;B}-\Box\dif_B\dif_CK^C_{\;\;A} \right) 
-6\Box(\dif_AV_B(y)+\dif_BV_A(y))+\nn\\&&+4\eta_{AB}\Box\dif_CV^C(y)+8\dif_A\dif_B\dif_CV^C(y)
+y_A S'_B + y_B S'_A = 0, \nn \\&&S'_A(y)=S_A(y)+3\Box^2V_A(y)-3\Box\dif_A\dif_CV^C(y).
\e{Heom2}
($S'_A$ is as arbitrary as $S_A$.) We may now perform a reduction using \rl{Hh} and \rl{ydotH2}. We find then 
the equations
\be
&& 3 \square_x \square_x h_{\mu\nu} - 3 \partial_\mu \partial^\lambda \square_x h_{\lambda \nu} - 3 \partial_\nu \partial^\lambda \square_x h_{\lambda \mu} + 2 \partial_\mu \partial_\nu \partial^{\rho} \partial^{\sigma} h_{\rho \sigma} + {} \nn \\[2mm]
&& \quad {} + \partial_\mu \partial_\nu \square_x h^\lambda_{\phantom{\lambda}\lambda} - \eta_{\mu\nu} \left( \square_x \square_x h^\lambda_{\phantom{\lambda}\lambda} - \partial^{\rho} \partial^{\sigma} \square_x h_{\rho \sigma} \right) = 0,
\e{conf4d}
which is perfectly consistent with the reduction of the Lagrangian in \rl{c04}. \rl{conf4d} are the equations for linearized conformal gravity in $d=4$. Notice that \rl{unphysical} and \rl{ydotH2} tell us that the only relevant physical field is $h_{\mu\nu}(x)$.

  \setcounter{equation}{0}
\section{Relativistic particle model for arbitrary spins} 
The O(N)-extended supersymmetric Lagrangian for relativistic particles of spin N/2 in $d=4$ is given by \cite{Marnelius:1988ab} (see also \cite{Gershun:1979fb,Howe:1988ft}) ($k,l=1,\ldots,N$)
\be
&&L={1\over 2v}\dox^2-\half i\psi_k\cdot\dot{\psi}_k-i{\rho_k\over v}\psi_k\cdot\dox-\half if_{kl}\psi_k\cdot\psi_l,
\e{a01}
where $\rho_k$, $f_{kl}$ together with the einbein variable $v$ are Lagrange multipliers. It yields the first class constraints
\be
&&p^2=0, \quad p\cdot\psi_k=0,\quad \psi_k\cdot\psi_l=0.
\e{a02}
The theory \rl{a01} is therefore a gauge theory. A BRST quantization is possible and was performed in \cite{Marnelius:1988ab}.
The quantization produces naturally a multispinor representation of a spin $N/2$ particle. However, spinor-tensor formulations are also possible to obtain. For integer spins ($N$ even) we get directly tensor fields if we introduce the oscillators \cite{Marnelius:1988ab}
\be
&&b_k^{\mu}\equiv{1\over\sqrt{2}}\bigl(\psi^{\mu}_{(2k-1)}-i\psi^{\mu}_{2k}\bigr),\quad k=1,\ldots,s\equiv N/2,
\e{a03}
which after quantization satisfy the anticommutation relations
\be
&&[b_k^{\mu}, b_l^{\nu\dagger}]_+=-\del_{kl}\eta^{\mu\nu}.
\e{a04}

We  apply now a Dirac quantization of the constraints in \rl{a02}. These constraints satisfy an $O(N)$ extended supersymmetric Lie algebra. In the following we consider only even $N$ and make use of the oscillator representation \rl{a03}.  Furthermore, we generalize  the treatment for $d=4$ in \cite{Marnelius:1988ab} to arbitrary even dimensions $d$. The general ansatz for a quantum state is
 (the sum is over all possible values of $n_j$, $j=1,\ldots,s\equiv N/2$)
\be
&&|F\hb=\sum_{n_j}F_{\mu_1\cdots \mu_{n_1};\nu_1\cdots \nu_{n_2}; \rho_1\cdots \rho_{n_3}; \cdots    }(x)|0\hb^{\mu_1\cdots \mu_{n_1};\nu_1\cdots \nu_{n_2};\rho_1\cdots \rho_{n_3};\cdots},
\e{a05}
where 
\be
&&|0\hb^{\mu_1\cdots \mu_{n_1};\nu_1\cdots \nu_{n_2};\rho_1\cdots \rho_{n_3};\cdots}
\equiv b_1^{\mu_1\dagger}\cdots b_1^{\mu_{n_1}\dagger}b_2^{\nu_1\dagger}\cdots b_2^{\nu_{n_2}\dagger}b_3^{\rho_1\dagger}\cdots b_3^{\rho_{n_3}\dagger}\cdots |0\hb,\nn\\
&& b_j^{\mu}|0\hb=0,\quad p_{\mu}|0\hb=0.
\e{a06}
The first constraints in \rl{a02} lead to the quantum conditions
\be
&&p^2|F\hb=0\quad\Leftrightarrow\quad \Box F_{\mu_1\cdots \mu_{n_1};\nu_1\cdots \nu_{n_2}; \rho_1\cdots \rho_{n_3}; \cdots    }(x)=0,
\e{a09}
\be
&p\cdot b_j|F\hb=0\quad&\Leftrightarrow\quad \dif^{\mu}F_{\mu\mu_2\cdots \mu_{n_1};\nu_1\cdots \nu_{n_2}; \rho_1\cdots \rho_{n_3}; \cdots   }(x)=0,\nn\\
&&\quad \quad \dif^{\nu}F_{\mu_1\mu_2\cdots \mu_{n_1};\nu\nu_2\cdots \nu_{n_2}; \rho_1\cdots \rho_{n_3}; \cdots   }(x)=0,\quad {\rm etc},
\e{a10}
\be
&p\cdot b_j^{\dagger}|F\hb=0\quad&\Leftrightarrow\quad \dif_{[\mu_1}F_{\mu_2\cdots \mu_{n_1+1}];\nu_1\cdots \nu_{n_2}; \rho_1\cdots \rho_{n_3}; \cdots   }(x)=0,\quad  {\rm etc},
\e{a11}
and the last constraints in \rl{a02} yield the conditions 
\be
&(b_k^{\dagger}\cdot b_l-b_l\cdot b_k^{\dagger})|F\hb=0&\Leftrightarrow\quad (b_k^{\dagger}\cdot b_l-d/2\del_{kl})|F\hb=0,
\e{a07}
\be
&(b_k\cdot b_l-b_l\cdot b_k)|F\hb=0&\Leftrightarrow \quad  b_k\cdot b_l|F\hb=0,\nn\\
&(b_k^{\dagger}\cdot b_l^{\dagger}-b_l^{\dagger}\cdot b_k^{\dagger})|F\hb=0&\Leftrightarrow \quad b_k^{\dagger}\cdot b_l^{\dagger}|F\hb=0.
\e{a08}
The  condition \rl{a07} determines $n_j$ to be $d/2$ which requires that $d$ is even. It also means that an $s$ field ($s\equiv N/2$) 
is a tensor field with $sd/2$ indices consisting of $s$ groups of antisymmetric sets of $d/2$ indices, \ie we have
\be
&&F_{\mu_1\cdots\mu_{d/2};\nu_1\cdots\nu_{d/2};\rho_1\cdots\rho_{d/2};\cdots}(x).
\e{a12}
Furthermore, the  condition in \rl{a07}  implies that $F_{\cdots}$ is symmetric under  interchange of any two groups of antisymmetric indices. The  two conditions in \rl{a08} require that all possible contractions of $F_{\cdots}$ are zero, \ie we have 
\be
&&F_{\la\mu_2\cdots\mu_n;\;\;\;\nu_2\cdots\nu_n;\rho_1\cdots\rho_n;\cdots}^{\quad\quad\quad\la}(x)=0,\quad {\rm etc}.
\e{a13}
Further details for $d=4$ are given in  \cite{Marnelius:1988ab}.  Below we give some examples.

\subsection{$s=1$ in $d=2$}
Here \rl{a11}-\rl{a07} yield
\be
&&\dif^{\mu}F_{\mu}(x)=0,\quad \Box F_{\mu}(x)=0,\nn\\
&&F_{\mu\nu}(x)=\dif_{\mu}\phi(x).
\e{a131}
Thus, here we have a spinless particle described by the scalar $\phi(x)$. It satisfies the Klein-Gordon equation due to the first relation. The second relation is then implied by the first and third relations. 

\subsection{$s=1$ in $d=4$ (spin one) \protect\cite{Marnelius:1988ab}}
Here \rl{a11}-\rl{a07} yield the standard free spin one field
\be
&&\dif^{\mu}F_{\mu\nu}(x)=0,\quad \Box F_{\mu\nu}(x)=0,\nn\\
&&F_{\mu\nu}(x)=\dif_{\mu}A_{\nu}(x)-\dif_{\nu}A_{\mu}(x).
\e{a14}

\subsection{$s=1$ in $d=6$}
Here \rl{a11}-\rl{a07} yield
\be
&&\dif^{\mu}F_{\mu\nu\rho}(x)=0,\quad \Box F_{\mu\nu\rho}(x)=0,\nn\\
&&F_{\mu\nu\rho}(x)=\dif_{\mu}A_{\nu\rho}(x)+\dif_{\nu}A_{\rho\mu}(x)+\dif_{\rho}A_{\rho\mu}(x),
\e{a15}
where $A_{\mu\nu}$ is antisymmetric (a two-form field).

\subsection{Spin two ($s=2$ in $d=4$) \protect\cite{Marnelius:1988ab}}
Here we find $F_{\mu\nu;\rho\la}=R_{\mu\nu\rho\la}$, the Riemann tensor: All properties of $R_{\mu\nu\rho\la}$ follow from \rl{a11}-\rl{a07}.

\end{appendix}

\bibliographystyle{utphysmod2}
\bibliography{biblio}
\end{document}